\documentclass[journal]{IEEEtran}
\usepackage[nocompress]{cite}
\usepackage{amsmath,amssymb,amsfonts}
\usepackage{algorithm}%
\usepackage{graphicx}
\usepackage{algpseudocode}

\usepackage{listings}%
\usepackage{subfig}
\usepackage{textcomp}
\usepackage{tabularx}
\usepackage{url}
\def\BibTeX{{\rm B\kern-.05em{\sc i\kern-.025em b}\kern-.08em
    T\kern-.1667em\lower.7ex\hbox{E}\kern-.125emX}}

\begin{document}

\title{Study of Cluster-Based Routing Based on Machine Learning for UAV Networks in 6G}

\author{Luis Antonio L. F. da Costa, Rodrigo C. de Lamare, Rafael Kunst, and Edison Pignaton de Freitas 

\thanks{Mr. da Costa and Dr. Kunst are with the University of Vale do Rio dos Sinos (UNISINOS), Brazil (e-mails: lalfcosta/rafaekkunst@unisinos.br), Dr. de Lamare is with  the Pontifical Catholic University - Rio de Janeiro (PUC-RIO), Brazil (e-mail: delamare@puc-rio.br) and Dr. Freitas is with the Federal University of Rio Grande do Sul, Brazil (email: edison.pignaton@inf.ufrgs.br)}}

\maketitle

\begin{abstract}
The sixth generation (6G) wireless networks are envisioned to deliver ultra-low latency, massive connectivity, and high data rates, enabling advanced applications such as autonomous {unmaned aerial vehicles (UAV)} swarms and aerial edge computing. However, realizing this vision in Flying Ad Hoc Networks (FANETs) requires intelligent and adaptive clustering mechanisms to ensure efficient routing and resource utilization. This paper proposes a novel machine learning-driven framework for dynamic cluster formation and cluster head selection in 6G-enabled FANETs. The system leverages mobility prediction using {Extreme Gradient Boosting (XGBoost)} and a composite optimization strategy based on signal strength and spatial proximity to identify optimal cluster heads. To evaluate the proposed method, comprehensive simulations were conducted in both centralized (5G) and decentralized (6G) topologies using realistic video traffic patterns. Results show that the proposed model achieves significant improvements in delay, jitter, and throughput in decentralized scenarios. These findings demonstrate the potential of combining machine learning with clustering techniques to enhance scalability, stability, and performance in next-generation aerial networks.
\end{abstract}

\begin{IEEEkeywords}
Computer Networks, Wireless Networks, Machine Learning, Network Clustering, Network Optimization
\end{IEEEkeywords}

\section{Introduction}
\label{sec:intro}

The sixth generation (6G) of wireless networks \cite{jpba} is envisioned to provide several new applications that focus on ultra-low latency, massive connectivity, and higher data rates. Breakthroughs in electronics, sensors, and communication technologies have paved the way for the development of compact unmanned aerial vehicles (UAVs). However, the capabilities of a single UAV may be limited, making the deployment of multiple UAVs essential to enhance performance and build more sophisticated systems. A Flying Ad Hoc Network (FANET) consists of a group of small interconnected UAVs that collaborate as a team to accomplish complex tasks. \cite{cappello2023using}.

Dense and complex environments, such as urban areas, stadiums, and large-scale industrial zones, present unique challenges for FANETs network performance due to their complex topographies and high user densities. These 6G communication environments are characterized by significant signal interference, heterogeneous propagation conditions, and dynamic user behavior, all of which can severely impact network efficiency and reliability \cite{pasandideh2023systematic}.

In order to attend the necessary requirements of these applications, FANETs must ensure communication stability and have high scalability. Recently, researchers have used clustering techniques to address routing problems in FANETs \cite{Abdulhae2022ClusterBasedRP,cbmlrout,rpreccf,cesg,rrs,rscf,rracf,iddllr,oclidd}. Such process consists in categorizing the network into small groups called clusters, while usually, each cluster incorporates cluster head nodes (CHs) and cluster member nodes (CMs) \cite{Hosseinzadeh2023ANF}. The most important step in network clustering (or clustering protocol) is the selection of the clusters heads, since the CH is responsible for managing the cluster and establishing inter-cluster and intra-cluster communication \cite{Lnsk2022ReinforcementLR}. 

This network optimization technique can solve some problems such as remote communication, scalability, and network reliability, as well as providing reliable and eﬃcient paths by reducing communication overhead when choosing routes and transferring data packets. Therefore, key aspects such as choosing the best CHs, managing network topology, and the choice of the routing protocol for the CMs, are essential steps to improving network eﬃciency in the clustering process for FANETs \cite{debasis2023energy}.

However, changes in the position of CMs can affect the cluster topology. Hence, cluster construction and management are very challenging and difficult due to frequent changes in FANETs. Understanding and addressing these challenges is critical for the successful deployment and operation of 6G networks \cite{asaamoning2021dynamic}. Therefore, designing an intelligent system for network clustering is essential to improve network performance in FANETs, especially in the context of centralized and decentralized communication environments. Machine learning methods are able to analyze the network topology and the cluster head selection may be optimized to consider key network parameters, all in order to construct an efficient system model that can ensure high performance and reliability on the operation of future-generation wireless networks \cite{abdulhae2023reinforcement}.

{Another important aspect to consider when dealing with FANETs is the scalability and performance trade-offs of these types of networks using advanced optimization and learning techniques. For example, in} \cite{xu2024multi} {the authors proposed a DRL-based fairness-oriented design for multi-UAV assisted mixed FSO/RF communication, while the work of} \cite{chen2024joint} {studied fairness and efficiency in CSMA/CA-based UAV MIMO ad hoc networks. In }\cite{li2024ground} {the authors focused on sub-terahertz channel modeling for ground-to-UAV links, and the work of} \cite{xu2025blockchain} {investigated blockchain-based AR offloading in UAV-enabled MEC networks. These contributions underline the increasing demand for scalable and intelligent UAV networking solutions, which motivates our ML-assisted clustering approach tailored for 6G FANETs.}

In this study, we propose a novel framework that utilizes machine learning to dynamically optimize cluster formation in 6G networks by analyzing user mobility patterns to predict optimal clustering configurations. Through comprehensive simulations, we show that our approach effectively reduces communication delay and jitter while achieving high data throughput in both centralized and decentralized network topologies, applicable to 5G and 6G environments, respectively. The findings also underscore the potential of deep learning to improve the adaptability and resilience of 6G networks, opening new avenues for future research and innovation in this evolving domain.

The remainder of this paper is organized as follows. Section \ref{sec:background} reviews essential background about network clustering in FANETs, as well as the relevant literature in this area. The proposed system model is presented in section \ref{sec:system}. The simulation scenario, performed experiments, and obtained results are presented and discussed in \ref{sec:results}. Finally, concluding the paper, Section \ref{sec:conclusion} presents final remarks and directions for future investigations.

\section{Background}
\label{sec:background}

The advent of sixth-generation (6G) wireless networks has introduced new paradigms in wireless communication, characterized by ultra-low latency, high reliability, and massive connectivity. These ambitious requirements pose significant challenges, particularly in dynamic and highly mobile environments such as Flying Ad Hoc Networks (FANETs) \cite{noman2023machine}. In this section, we explore existing research on cluster formation strategies in FANETs and the application of machine learning techniques for network optimization, with a focus on identifying key limitations and motivating the need for a novel deep learning-based clustering framework.

Clustering is a widely adopted strategy to manage the scalability and reliability of FANETs. It involves grouping UAV nodes into clusters to reduce communication overhead, optimize routing, and balance resource utilization. Traditional clustering approaches in FANETs are often derived from Mobile Ad Hoc Network (MANET) protocols, such as:

\begin{itemize}
    \item \textbf{LEACH (Low-Energy Adaptive Clustering Hierarchy):} Originally designed for sensor networks, LEACH selects cluster heads based on probabilistic models to minimize energy usage. However, its performance degrades in high-mobility scenarios typical of FANETs \cite{sefati2022cluster}.
    \item \textbf{Weighted Clustering Algorithms (WCA):} These use metrics such as node degree, mobility, and battery power to select cluster heads. Although more adaptive, WCA-based methods struggle to keep up with the fast topology changes in FANETs \cite{wang2024fast}.
    \item \textbf{Mobility-aware clustering:} Some recent research, such as the work of \cite{karpagalakshmi2024energy}, have proposed bio-inspired mobility prediction models to improve cluster stability. While these methods show improved performance over static algorithms, they often rely on simplistic mobility models or lack real-time adaptability.
\end{itemize}

Nevertheless, existing clustering methods are limited by their reliance on heuristic or rule-based decision-making, which does not scale well in dynamic and high-density 6G environments. Moreover, most of them fail to leverage the vast amount of network and mobility data available in modern UAV networks.

{Recent works have increasingly explored the integration of UAV-assisted communications, advanced wireless access technologies, and AI-driven optimization techniques to enhance the performance of next-generation wireless networks under strict resource and energy constraints. For instance, the work of }\cite{zhang2023capacity} {proposed a Double Deep Q-Network (DDQN)-based framework that jointly optimizes UAV trajectory and RIS phase shift design in RIS-UAV-assisted NOMA networks. Their approach maximizes system capacity while accounting for UAV energy limitations, demonstrating significant improvements in network scalability and robustness, particularly in dynamic or emergency scenarios.}

{In the context of content delivery, the work of} \cite{zhou2024fd3pg} {presented a Federated Distributed Deep Reinforcement Learning method (FD3PG) for recommendation-enabled edge caching in multi-tier edge-cloud networks. The authors employ single-agent DDPG extended to a federated multi-agent setting, enabling personalization and avoiding local optima. Their results show that FD3PG substantially reduces delivery delay and improves cache hit rates compared to existing baselines, highlighting the potential of distributed learning for edge intelligence.}

{Complementarily, the work of} \cite{zhou2025qos} {investigated a QoS-oriented framework for NOMA-enhanced UAV-assisted MEC systems, focusing on balancing task delay and energy consumption in highly dynamic environments. They introduced the System Overhead Ratio (SOR) as a metric for capturing this trade-off and proposed a Lyapunov-based low-complexity online method (LORT) to jointly optimize resource allocation, transmission power, and UAV trajectory. Simulation results confirm that LORT reduces SOR by 10–25\% as compared to benchmarks, demonstrating its adaptability for real-world UAV-MEC scenarios.}

The use of machine learning (ML) in wireless networks has grown significantly, particularly in areas such as spectrum allocation, traffic prediction, and anomaly detection. For FANETs, ML methods offer the potential to adaptively manage complex interactions between mobility, channel conditions, and user demands.

\begin{itemize}
    \item \textbf{Reinforcement Learning (RL):} RL has been applied to optimize routing paths and transmission strategies in UAV networks. Works such as \cite{tang2024deep} demonstrate that RL can significantly reduce latency in UAV swarms by learning optimal communication patterns. However, RL often requires extensive exploration, which can be impractical in latency-sensitive applications.
    \item \textbf{Supervised Learning for Link Prediction and Clustering:} Some studies \cite{prakash2024reinforcement} have employed supervised learning to predict link failures or optimize cluster membership. These methods require large labeled datasets and often do not generalize well to unseen topologies.
    \item \textbf{Unsupervised and Semi-supervised Learning:} These approaches have been explored for anomaly detection and traffic classification \cite{jaiswal2024comparative}, but their application in real-time clustering remains limited.
\end{itemize}

While promising, many ML approaches suffer from a lack of integration between mobility, traffic, and network data. Most existing solutions focus on isolated problems rather than offering a holistic view of network optimization. Furthermore, few models are designed to operate in a real-time, distributed, and adaptive manner suitable for 6G FANETs.

Deep learning, with its powerful feature extraction capabilities, has recently been explored for clustering problems in wireless networks:

\begin{itemize}
    \item \textbf{Deep Reinforcement Learning (DRL):} these approaches have shown promise in dynamic environments \cite{li2024reinforcement}, where agents learn optimal policies for node association and routing. However, the complexity and convergence time of DRL can hinder practical deployment.
    \item \textbf{Graph Neural Networks (GNNs):} Some cutting-edge works \cite{song2025gnnppor} have proposed using GNNs to model the network topology and derive optimal cluster structures. Although GNNs offer a strong representation of node relationships, they often require high computational resources and are still in early development stages for FANET applications.
    \item \textbf{Hybrid DL Models:} Few studies \cite{priyadharshini2024efficient} have proposed hybrid architectures that combine CNNs or RNNs with contextual data such as node trajectories and QoS metrics to predict network states and guide cluster formation.
\end{itemize}

Despite these developments, no unified deep learning framework currently exists that integrates mobility patterns, traffic demands, and dynamic topology to make real-time clustering decisions in FANETs under 6G standards. There remains a need for models that can learn from heterogeneous data, adapt quickly to network changes, and be deployed with manageable complexity.

{Beyond performance optimization, it is worth mentioning that recent studies have also emphasized the importance of reliability and fairness in UAV and 6G networks. For instance,} \cite{luo2025convergence} {explored the convergence of symbiotic communications and blockchain to achieve sustainable and trustworthy 6G architectures, highlighting the role of security and reliability guarantees. The work of} \cite{9286738} {proposed dynamic network function provisioning through “network-in-a-box” concepts for industrial applications, addressing scalability and dependability in mission-critical settings. In} \cite{10714036} {the authors introduced proportional fairness-aware scheduling in space–air–ground integrated networks, ensuring balanced resource allocation under dynamic conditions. Similarly,} \cite{chen2024and} {proposed the Lasagna air–ground integrated infrastructure design, which provides robust connectivity layers tailored to future safety-critical services. These works underscore that, in addition to average performance gains, robustness and worst-case guarantees are essential, which motivates our proposed ML-assisted clustering framework as a step toward reliable 6G UAV networking.}

The research gaps of these several fields of study motivate the need for a novel deep learning-based clustering framework capable of dynamically optimizing cluster configurations in 6G FANETs. By leveraging the joint analysis of user mobility and traffic characteristics, such a framework can significantly enhance the efficiency, reliability, and adaptability of next-generation wireless networks.

\section{System}
\label{sec:system}

This section explores the design details of the proposed system model for network clustering, as well as the optimization problem developed for the cluster head selection. The application scenario of both network topologies is also presented: a centralized topology for 5G networks, and then a distributed topology for 6G networks.

\subsection{System Model}

\begin{figure}[!htbp]
\centering
\includegraphics[width=0.5\textwidth]{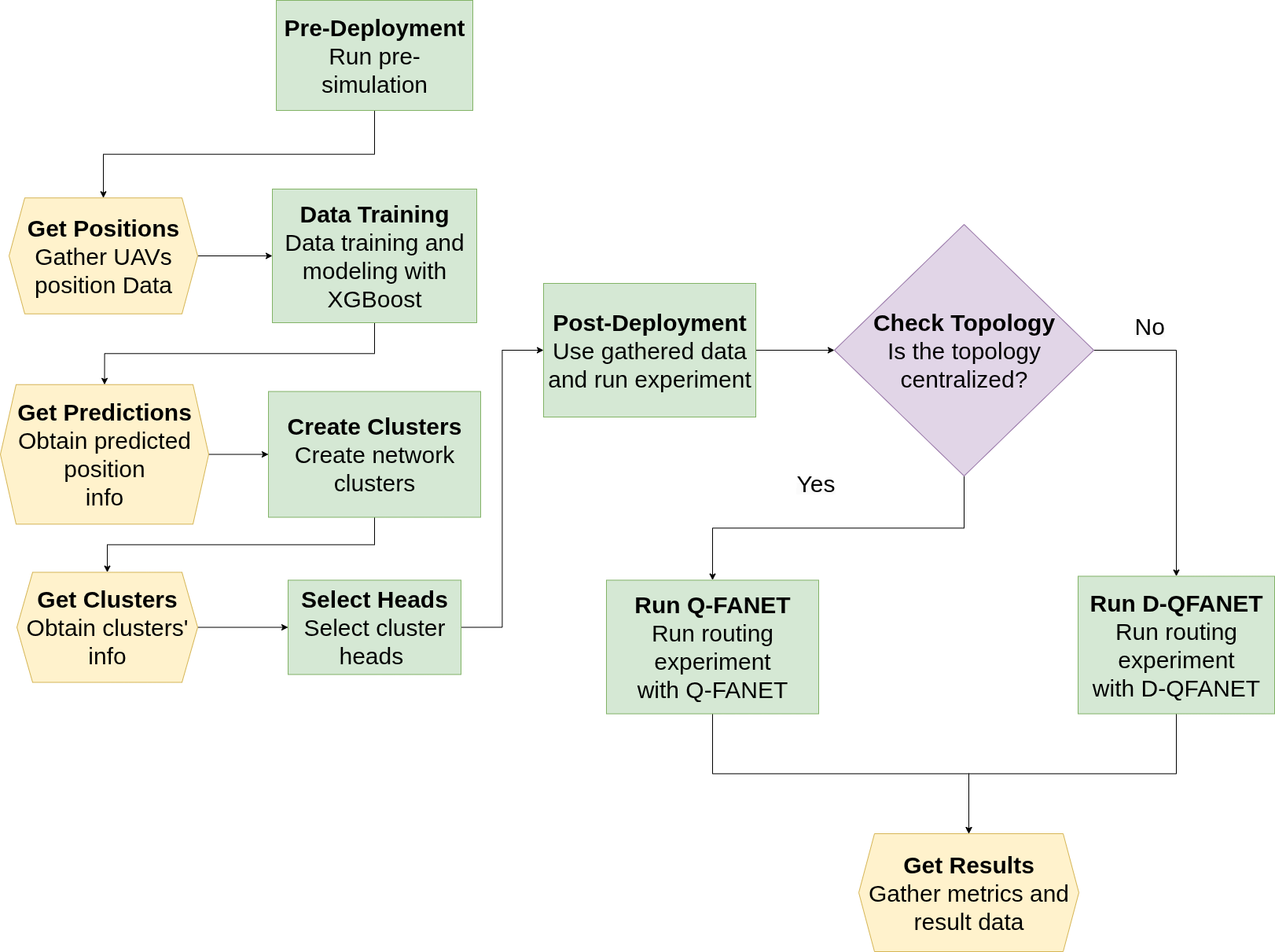}
\caption{System workflow for network clustering and routing simulation.}
\label{fig:cluster_system}
\end{figure}

In order to perform network clustering that uses machine learning approaches, a system model was envisioned following a set of steps and tasks. Figure \ref{fig:cluster_system} summarizes the workflow of our proposed approach for network clustering in order to create a suitable communication environment to test simulations in the 5G and 6G contexts.

This workflow is detailed as it follows:
\begin{itemize}
    \item {\textbf{Pre-Deployment:} In order to obtain a position dataset for the mobile stations, a pre-simulation of a FANET with mobility is run;}
    \item {\textbf{Get Positions:} Gather UAVs position data by recording and storing the mobile station's position history;}
    \item {\textbf{Data Training:} The dataset is separated in two other sets (training and testing) and \textit{XGBoost}} \cite{Chen2016XGBoostAS} {is applied to train and generate a model;}
    \item {\textbf{Get Predictions:} Obtain predicted position info from the \textit{XGBoost} model that creates a dataset with predicted positions for each mobile station;}
    \item {\textbf{Create Clusters:} The network clusters are created using the Elbow method} \cite{syakur2018integration} {and the Knee Point method} \cite{zhao2008knee} {to automate the elbow method by mathematically identifying the point of maximum curvature, enhancing objectivity and reproducibility;}
    \item {\textbf{Get Clusters:} The clusters' info is recorded and stored;}
    \item {\textbf{Select Heads:} A Mixed-Integer Linear Programming (MILP) problem is suggested to select the clusters heads depending on their distance and signal power to the other mobile stations withing the cluster;}
    \item {\textbf{Post-Deployment:} The gathered data for the clusters and their respective heads is used to run new network experiments;}
    \item {\textbf{Check Topology:} Centralized or Decentralized?}
        \item {\textbf{Run Q-FANET:} If the topology is centralized, the network simulation will use Q-FANET} \cite{Costa2021QFANETIQ}   {as the main routing mechanism;}
        \item {\textbf{Run D-QFANET:} If the topology is decentralized, the network simulation will use D-QFANET} \cite{daCosta2024RoutingPE}   {as the main routing mechanism;}
    \item {\textbf{Get Results:} Finally, specific measurements are recorded during the experiment and selected metrics are analyzed from the resulting data.}
\end{itemize}

For the training of the \textit{XGBoost} model to predict the final position of the mobile stations, a set of parameters has been chosen:

\begin{itemize}
    \item \textbf{Loss function:} Squared Error
    \item \textbf{Evaluation metric:} Root Mean Squared Error
    \item \textbf{Decision Tree maximum depth:} 6
    \item \textbf{Learning rate:} 0.1
    \item \textbf{Fraction of features to be randomly sampled for constructing each tree:} 1
    \item \textbf{Fraction of the training data to be randomly sampled for growing each tree:} 1
\end{itemize}

These parameters strike a balance between performance and preventing overfitting. They are commonly used in regression tasks like predicting final positions of elements in a grid based on input features \cite{Chai2014RootMS}.

\subsection{Network Model}

Standard FANET’s communication architecture is divided into two categories based on its connectivity, such as centralized and decentralized (distributed). In 5G networks, for example, a centralized network topology is necessary to handle complex interactions between various network components and efficiently manage network resources \cite{alba2021enabling}.

\begin{figure}[h!]
    \centering
    \subfloat[A centralized network topology design in 5G Core Networks.]{
        \includegraphics[width=0.45\textwidth]{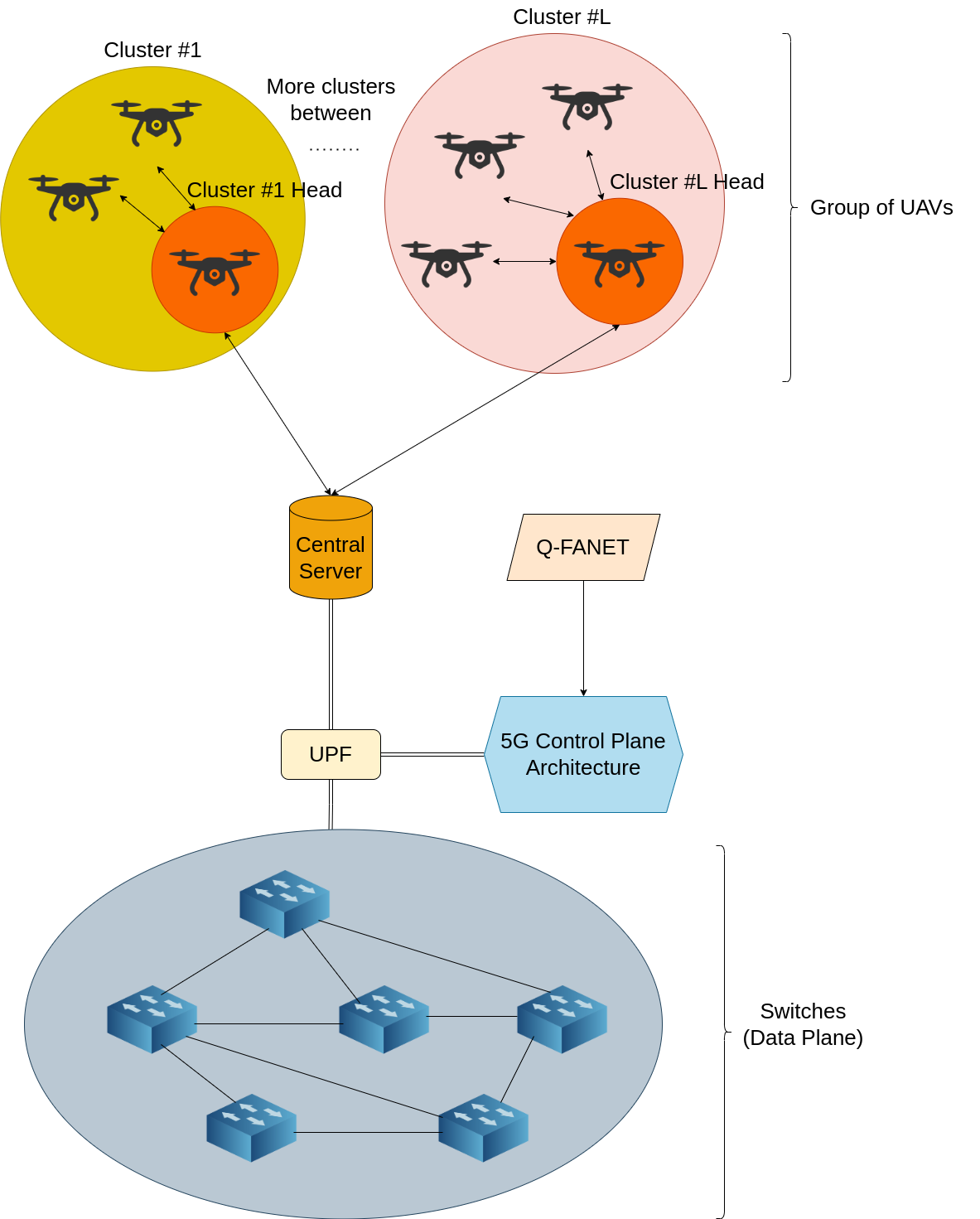}
        \label{fig:cnt_topo}
    }
    \hspace{0.05\textwidth} 
    \subfloat[A distributed network topology design in 6G Core Networks.]{
        \includegraphics[width=0.45\textwidth]{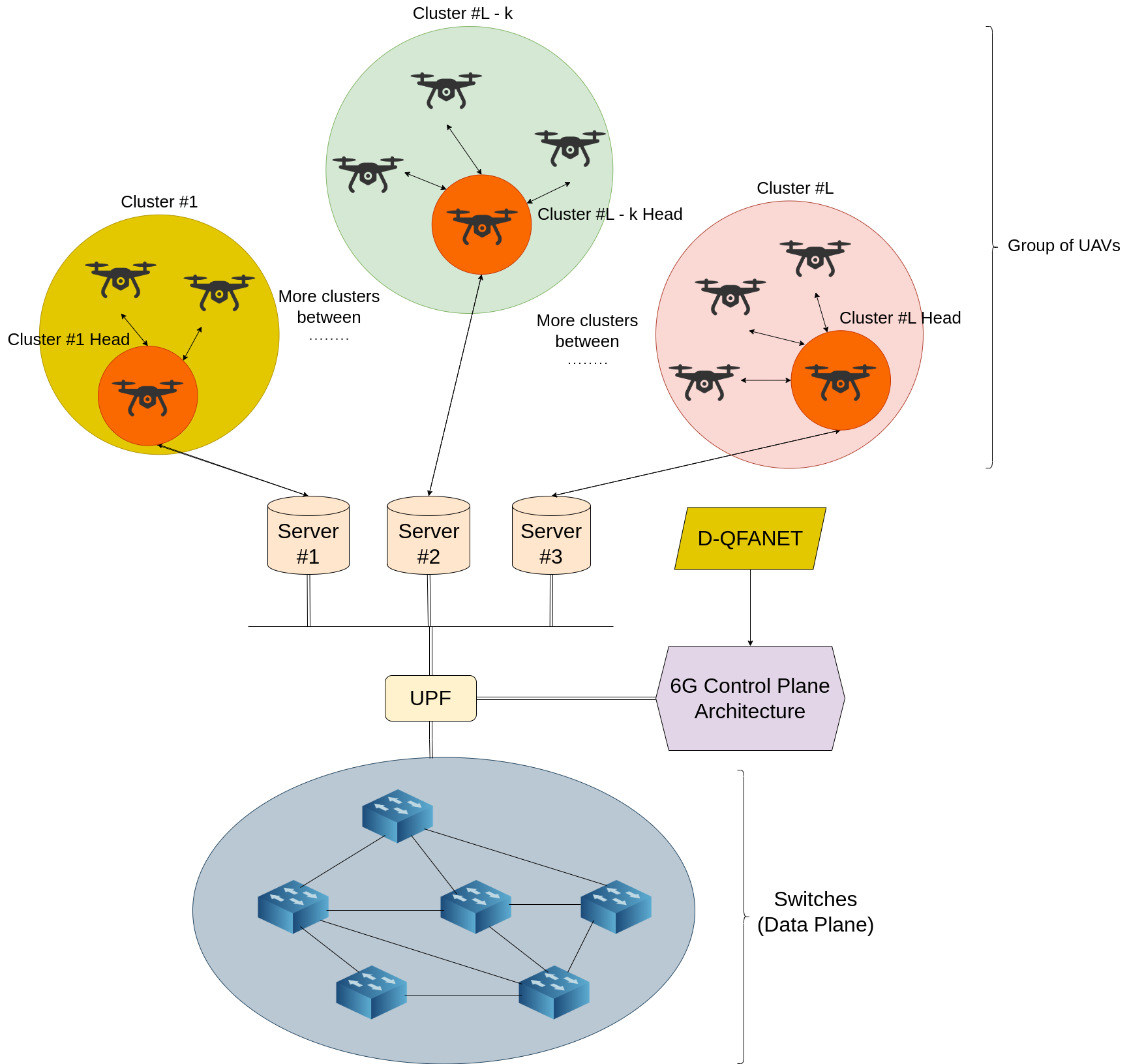}
        \label{fig:dst_topo}
    }
    \caption{Network topology designs.}
    \label{fig:sidebyside}
\end{figure}

Figure \ref{fig:cnt_topo} details the centralized network topology proposed for the 5G network architecture, with our approach creating the clusters and selecting the appropriate cluster heads. It comprises a centralized server connected to a group of several UAVs. The server is then connected to the UPF (User Plane Function) module that connects to the Data Network in the Data Plane. This module also connects directly to the 5G core network, which consists of several key components/modules, each serving a specific purpose in facilitating advanced telecommunications services. The Q-FANET routing algorithm feeds the routing information through the 5G core architecture and passes it to the network. Finally, the Data Plane is composed of a network of several connected switches that are responsible for handling the routing of large amounts of data, if necessary.

However, the current 5G architecture does not actually support network intelligence. Hence, one of the goals of the future 6G core networks is to provide network intelligence to manage its network service function and the element of the architecture. For this reason, a new module is present in the 6G core network, the Network Data Analytics Function (NWDAF) module. NWDAF is responsible for collecting, analyzing, and providing network data analytics. Therefore, with the generated routing information, the NWDAF module determines the routing policy for the controller, which can then establish the optimal routing scheme for the distributed topology.

In Figure \ref{fig:dst_topo} are described the details of the distributed network topology proposed for the 6G network architecture. The key difference from the centralized topology is that in the distributed one, there are three servers, each connected to different groups of several UAVs. Moreover, in this scenario the D-QFANET is the routing algorithm that feeds routing information through the 6G control plane architecture and passing it to the network.

For both the network topologies, the network clustering approach is able to eliminate the necessity of Access Points, more specifically, RAN (Radio Access Network) antennas, and NG-RAN (Next-gen Radio Access Network) antennas, for 5G and 6G, respectively. In this manner, the cluster heads become the main communication receptors that will forward information packets from their respective UAV clusters to the servers, and backwards, creating an ad-hoc or infrastructureless network, where devices communicate directly with each other.

\subsection{Cluster head selection}

The main objective of the network clustering is to create a network environment that offers the lower communication delay possible between the mobile stations and host servers. Hence, the goal is to minimize an objective function, \textit{e.g}, the chosen cluster head should minimize the weighted sum of distances and maximize the signal power to all other stations within the same cluster, while satisfying the constraint that there is exactly one cluster head per cluster. This problem is mathematically expressed as follows. 

\begin{equation}
   \begin{split}
    \min_{w, x_i} & \sum_{l=1}^L(\sum_{i=1}^{M}\sum_{j=1,j \neq i}^{M}d_{ij}x_i - w\sum_{i=1}^{M}\sum_{j=1,j \neq i}^{M}p_{ij}x_i) \\
    {\rm subject ~to} & \sum_{i=1}^{M}x_i=1, 
     \end{split}
     \label{eq:min_prob}
\end{equation}

where the variables are defined as:
\begin{itemize}
    \item $x_i$: Binary variable indicating whether station $i$ is chosen as the cluster head (1 if chosen, 0 otherwise)
    \item $d_{ij}$: Distance between station $i$ and station $j$ within the same cluster
    \item $p_{ij}$: Signal power of station $i$ in relation to station $j$ within the same cluster
    \item $w$: Weight factor that balances the importance of distance and signal power.
\end{itemize}
  
The optimization problem formulated in \eqref{eq:min_prob}, can be categorized as a combinatorial optimization problem, \textit{e.g.}, finding the optimal solution among a countable, possibly finite, set of options. Such classification is justified by the discrete nature of the decision variables being binary (assuming values 0 or 1), representing the choice of stations as cluster heads. Similar problems have been reported in the literature for detection \cite{spa}, estimation \cite{jidf}, resource allocation \cite{rsprec} and distributed processing \cite{lrcc}.  

{Furthermore, the optimization objective function for choosing a cluster head within each cluster is written as:}
\begin{equation}
J_i(w) = \sum_{j \neq i} \Big( \underbrace{\tilde{d}_{ij}}_{\text{closer is better}} 
\;-\; w \cdot \underbrace{\tilde{p}_{ij}}_{\text{stronger is better}} \Big)
\end{equation}
  {where $w \in [0,1]$ controls the trade-off between distance and power. Here, $\tilde{d}_{ij}$ and $\tilde{p}_{ij}$ are normalized values of the inter-station distance and signal power, respectively, ensuring that both are dimensionless and comparable. Therefore, the parameter $w$ can be interpreted as:}

\begin{itemize}
    \item   {$w=0$: the cluster head is chosen solely by geometric compactness, i.e., the station closest on average to all others.}
    \item   {$w=1$: the cluster head is chosen solely by \textbf{signal power}, i.e., the station that maximizes link quality to all others.}
    \item   {$0<w<1$: an explicit trade-off between distance minimization and power maximization.}
\end{itemize}

  {To illustrate the impact of the weight parameter $w$, we evaluated a cluster with five stations ($s_1, s_2, s_3, s_4, s_5$). The objective $J_i(w)$ was computed for each candidate head across $w \in [0,1]$, after min--max normalization of distance and power values.}

\begin{figure}[!htbp]
\centering
\includegraphics[width=0.45\textwidth]{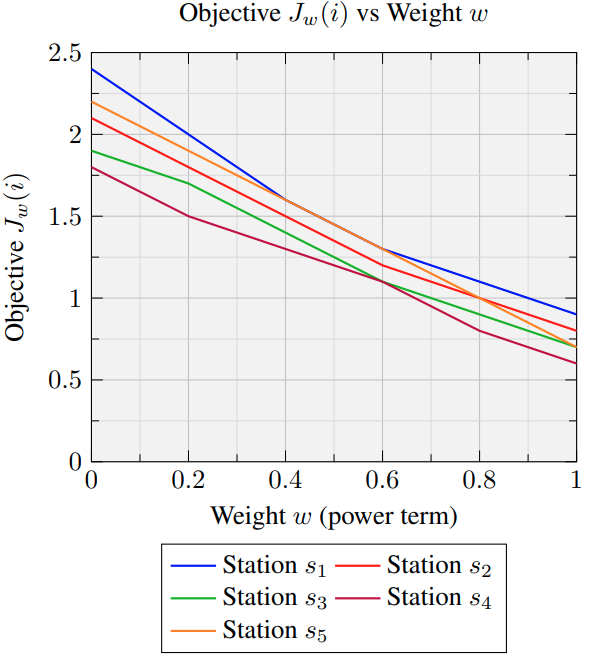}
\caption{Objective function $J_i(w)$ for each candidate head as a function of the weight $w$.}
\label{fig:objective-weight}
\end{figure}

  {As shown in Figure} \ref{fig:objective-weight},   {station $s_4$ consistently achieves the lowest value of the objective function across the entire range of $w$. This indicates that $s_4$ is the optimal cluster head regardless of how much emphasis is placed on distance versus power. In contrast, other stations either remain dominated throughout or become competitive only in very narrow ranges of $w$.}

Furthermore, the search for the best cluster configuration, among a vast set of possibilities, and the presence of restrictions, corroborate this categorization: the number of possible combinations of stations as cluster heads grows exponentially with the total number of stations, which makes the search space significantly large. The combinatorial complexity of this problem suggests the application of heuristic methods, such as genetic algorithms and particle swarm optimization, to find good-quality solutions. Thus, the mathematical expression in question provides a means of restricting each station to be assigned to only one cluster.

Such optimization problem could be solved by heuristic methods like Genetic Algorithms \cite{zeebaree2017combination} or Particle Swarm Optimization Algorithms \cite{rana2011review}, which can be applied for larger or more complex scenarios in network clustering. However, for model simplification purposes, the solution to this problem is obtained by a composite score computed for each station that considers the minimal distance and the stronger signal power from the other mobile stations inside the cluster.

First, the Received Signal Strength Indicator (RSSI) between the candidate cluster head and all other stations in the cluster is measured. Then, the Euclidean distance between the candidate cluster head and all other stations is computed. Next, the score for each station is computed as the difference between the average signal strength and the average distance to other stations in the cluster, giving priority to stations that are closer to others while also having strong signal strength. The station with the highest score within each cluster is selected as the cluster head, ensuring that the selected cluster head has both a strong signal and is centrally located relative to the other stations within the cluster.

In practice, to efficiently select the cluster head within each cluster without solving the MILP formulation, a heuristic scoring function is adopted. For every candidate node \( i \) in a cluster, a score is computed as the difference between the average received signal strength from node \( i \) to all other nodes in the same cluster and the average Euclidean distance to those nodes:

\begin{equation}
\text{Score}_i = \frac{1}{M - 1} \sum_{\substack{j=1 \\ j \ne i}}^M p_{ij} - \frac{1}{M - 1} \sum_{\substack{j=1 \\ j \ne i}}^M d_{ij}
\end{equation}

where:
\begin{itemize}
  \item \( p_{ij} \) is the signal power (e.g., RSSI) received from candidate node \( i \) to node \( j \),
  \item \( d_{ij} \) is the Euclidean distance between node \( i \) and node \( j \),
  \item \( M \) is the number of nodes in the cluster.
\end{itemize}

The node with the highest score is selected as the cluster head:

\begin{equation}
i^* = \arg\max_{i \in \{1, \dots, M\}} \text{Score}_i
\end{equation}

This method ensures the selection of a cluster head that is both centrally located and maintains strong signal power to other members, aligning with the objective of minimizing intra-cluster communication delay.

Moreover, this solution will identify the station within each cluster that optimizes the trade-off between minimizing the distance to other stations and maximizing the signal power to them, resulting in the lowest possible communication delay within the cluster. 

\subsection{Complexity Analysis}

Analyzing the computational complexity of the cluster head selection algorithm is crucial for understanding its scalability and real-world applicability, particularly in dynamic and resource-constrained environments such as FANETs. By quantifying the time and space complexity of the selection process, researchers can better evaluate the algorithm's feasibility for deployment in large-scale networks or highly mobile scenarios \cite{Cui2022}. 

A clear complexity analysis allows for informed comparisons with alternative methods, such as exhaustive MILP-based optimization or metaheuristic approaches, and helps highlight trade-offs between accuracy and computational efficiency. Furthermore, it guides the development of optimizations or approximations that retain acceptable performance while reducing overhead. For other researchers, this analysis serves as a valuable benchmark and reference, enabling them to reproduce results, adapt the approach to different contexts, and build upon it with confidence in its efficiency and limitations.

Let $L$ be the number of clusters and $M$ be the average number of mobile stations per cluster. The algorithm iterates over each cluster and, for each station within a cluster, computes the total signal strength and Euclidean distance to every other station in the same cluster. These operations are used to evaluate a score for selecting the optimal cluster head.

Within a single cluster, each station is compared with every other station, resulting in a nested loop structure over M stations. Consequently, the computational complexity for processing one cluster is $O(M^2)$. As the algorithm performs this procedure for each of the $L$ clusters, the overall time complexity becomes: $O(L \cdot M^2)$.

The space complexity is comparatively low. In addition to storing the input cluster dictionary and the output cluster heads, the algorithm uses only a constant amount of temporary memory for each station's score, total signal strength, and total distance. Therefore, the space complexity is: $O(L)$.

In summary, the algorithm has quadratic time complexity with respect to the number of stations per cluster, which is efficient for small to moderately sized clusters. However, for scenarios involving a large number of stations within each cluster, optimizations or approximations to the algorithm may be required to ensure scalability.

\begin{figure}[!htbp]
\centering
\includegraphics[width=0.50\textwidth]{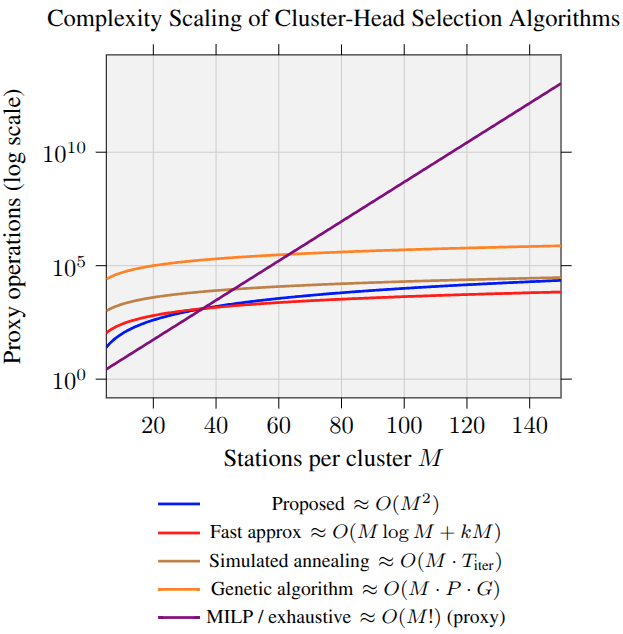}
\caption{Complexity of cluster-head selection \emph{per cluster} as $M$ grows (log-scale). The proposed pairwise signal-strength scoring scales quadratically; a fast spatial approximation using k-d trees + $k$NN trends toward $O(M\log M + kM)$. Metaheuristics are linear in $M$ but with large constants from iterations/populations, while MILP/exhaustive search grows super-exponentially.}
\label{fig:ch_complexity_per_cluster}
\end{figure}

{For the signal-strength-based CH selection considered in this work, the per-cluster cost is $O(M^2)$ due to pairwise aggregation of received power and distances among M stations, yielding an overall $O(L \dot M^2)$ across L clusters. Figure} \ref{fig:ch_complexity_per_cluster} {contrasts this with typical alternatives: exhaustive/MILP formulations (factorial growth), metaheuristics such as simulated annealing and genetic algorithms (linear in $M$ but with large constants from iterations and population sizes), and a fast spatial approximations that replaces all-pairs scoring with a k-d tree and k-nearest-neighbor neighborhood scoring, reducing the cost toward $O(MlogM +kM)$. These trends explain why our method is practical for small-moderate M.}

\section{Evaluation}
\label{sec:results}

In order to evaluate the performance of the proposed system model for network clustering with different network architectures and topologies, a set of experiments is proposed. This group of simulations is not yet integrated with the core architectures of both 5G and 6G networks and deal only with the centralized and decentralized topologies. In this manner, the Mininet Wi-Fi simulator \cite{Fontes2015MininetWiFiES} was used for the experiments, integrating the usage of Q-FANET and D-QFANET as routing protocols for the centralized and decentralized topologies, respectively.

{While it is true that network simulators such as ns-3 or hardware testbeds provide finer-grained fidelity, the choice of Mininet-WiFi in this study is deliberate and justified by its unique advantages. Mininet-WiFi is an emulator rather than a pure simulator, meaning that it executes real Linux networking stacks and protocol implementations instead of abstracted models. This enables experiments to capture practical effects such as queueing, routing, and interference interactions at the protocol level, while still running in a fully controllable and reproducible software environment.}

  {Compared to ns-3 simulator} \cite{henderson2008network},   {which is highly detailed but model-driven, Mininet-WiFi offers faster prototyping, seamless integration with real applications (since it runs unmodified binaries), and flexible support for emulating wireless mobility and topology changes. These features make it particularly suitable for evaluating relative performance trends (\textit{e.g.}, delay/jitter reduction across centralized vs. decentralized clustering) rather than absolute physical-layer accuracy.}

  {While real UAV testbeds would provide the highest fidelity, they also introduce significant practical barriers, such as hardware cost, flight safety constraints, and lack of scalability for large swarms. Mininet-WiFi therefore strikes a pragmatic balance: it enables repeatable, large-scale experiments that can validate algorithmic design choices and demonstrate end-to-end performance improvements under realistic protocol stacks. Following the methodology adopted in several prior works, our focus is not on reproducing exact wireless channel conditions but on evaluating the network-level behavior and benefits of the proposed clustering and routing framework.}

  {To mitigate limitations, we designed experiments that rely on comparative performance evaluation (centralized vs. decentralized, with and without clustering), where the relative differences are meaningful regardless of absolute delay/jitter accuracy. Future work, as noted in our conclusion, includes extending the study to ns-3 with 5G/6G modules and eventually to hardware-in-the-loop UAV swarms, which will further validate and complement the findings obtained in Mininet-WiFi.}

\subsection{Simulation setup}

The simulation is based on a a 3GPP video traffic model \cite{3GPPTraf68:online}, which involves generating packets based on a realistic video streaming pattern. The packet size and inter-arrival times are determined according to the video encoding (e.g., H.264/AVC). For this setup, is set an average packet size of 1024 bytes (with a variation of 256 bytes) and a average inter-arrival time of $30 ms$ based on a Poisson process. The packet size is generated in a normal distribution, as for the inter-arrival time is generated in an exponential distribution. Moreover, all the packets are sent through the network using UDP (User Datagram Protocol) sockets, in order to meet the stringent latency requirements of ultra-fast networks for 5G and 6G communication environments.

In the simulations, 100 UDP packets of various sizes are sent from each of the 25 mobile stations to their respective cluster heads, and to the central server (centralized topology) or to the distributed servers (distributed topology), for a period of $3600s$. In both scenarios, each UAV is assigned a random signal power between $60$ and $80$ $dBm$, which is one of the parameters considered for choosing a mobile station as a cluster head. Since both simulation scenarios will be run with the same clustering model, the network topologies will have the same number of clusters (3) and selected cluster heads. The complete set of parameters used for the experiments is detailed in Table \ref{tab:paremters}:

\begin{table}[!htbp]
\begin{center}
\caption{Simulation Parameters Setup.}
\begin{tabular}{ c c }
   ine \\[-1.8ex] 
 Parameters & Settings \\ 
   ine \\[-1.8ex] 
 Area size & 500m $\times$ 500m \\  
 Number of nodes & 25  \\
 Radio propagation & propagation range, range = 500m\\
 Interferences & interferences orthogonal\\
 Modulation & modulation bpsk\\
 Mobiliy Model & random waypoint\\
 Antenna & antenna omnidirectionnal\\
 Battery & energy linear\\
 HELLO Interval & 100ms\\
 Expire Time & 300ms\\
 Initial Q-Value & 0.0\\
 minspeed & 0 m/s\\
 maxspeed & 15 m/s\\
 UAV Signal Power & 60-80 dBm \\
 Data packet & 1024 Bytes (256 Bytes variation)\\
 SINR weight & 0.7\\
 Latency Threshold & 10ms \\
 Look back for Q-Noise+ (l) & 10\\
 $w$ & $0 < w < 1$\\
 $\alpha$ & 0.2\\
 $\epsilon$ & 0.2\\
   ine\\[-1.8ex]
\end{tabular}
\label{tab:paremters}
\end{center}
\end{table}

  {The choice of SINR weight (0.7) and $\alpha$ (0.2) follows conventions in prior works on UAV optimization and clustering} \cite{zhang2023capacity, zhou2025qos},   {where similar ranges have been adopted to balance conflicting metrics such as throughput, delay, and energy consumption. These values were also calibrated for numerical stability, and preliminary sensitivity analysis confirmed that the overall performance trends remain consistent. Similar weighting strategies can also be found in classic clustering frameworks such as LEACH} \cite{926982}.

\subsection{Results and Discussion}

Based on this set of parameters and the simulation setup, the main goal of the proposed experiments is to present the behavior of our proposed system model for network clustering in different network scenarios in terms of end-to-end delay, jitter, and data throughput.

\begin{figure}[h!]
    \centering
    \subfloat[Centralized topology (5G)]{
        \includegraphics[width=0.45\textwidth]{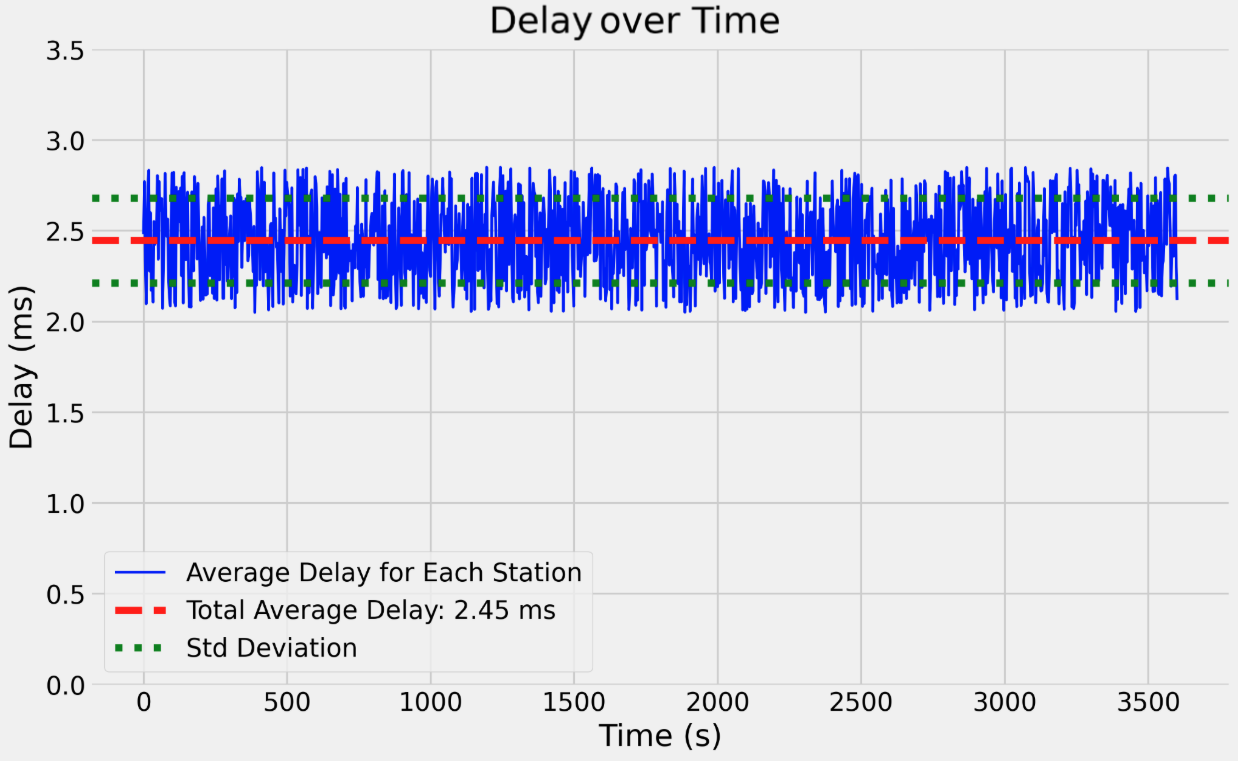}
    }
    \hspace{0.05\textwidth} 
    \subfloat[Decentralized topology (6G)]{
        \includegraphics[width=0.45\textwidth]{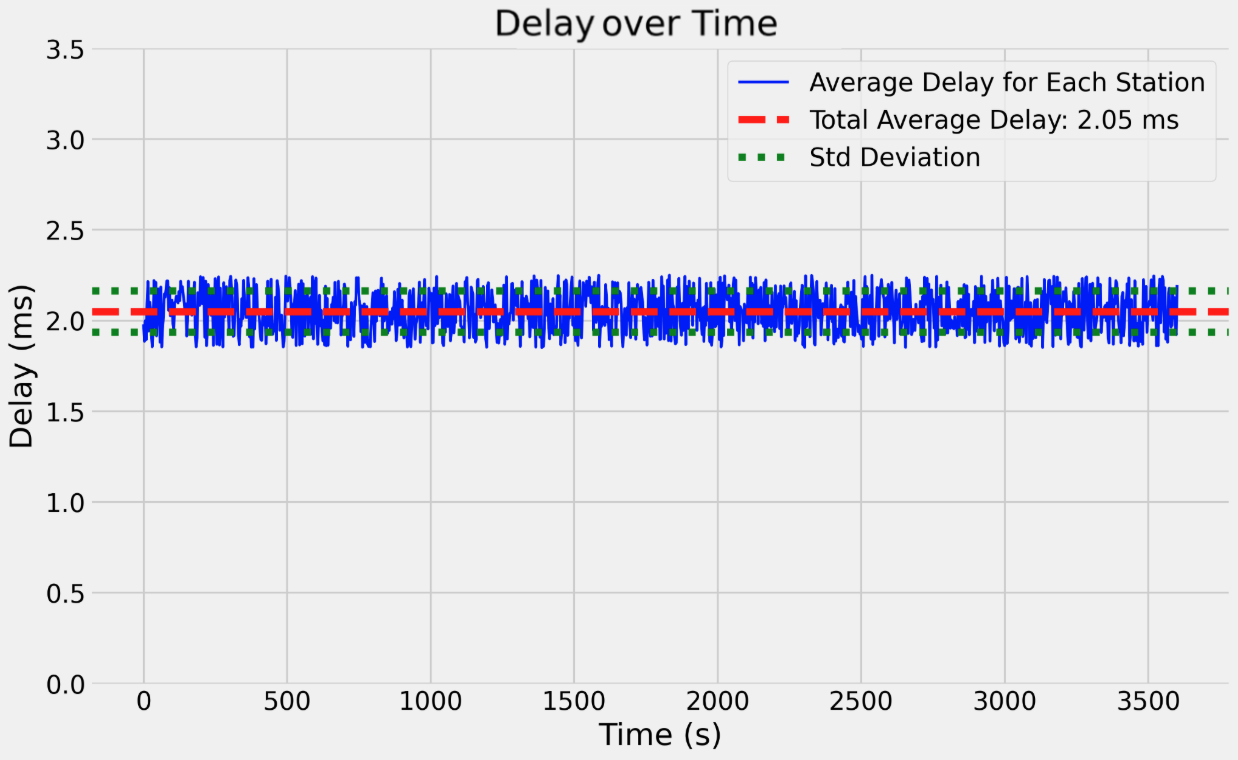}
    }
    \caption{Average delay for all stations.}
    \label{fig:delay}
\end{figure}

\begin{figure}[h!]
    \centering
    \subfloat[Centralized topology (5G)]{
        \includegraphics[width=0.45\textwidth]{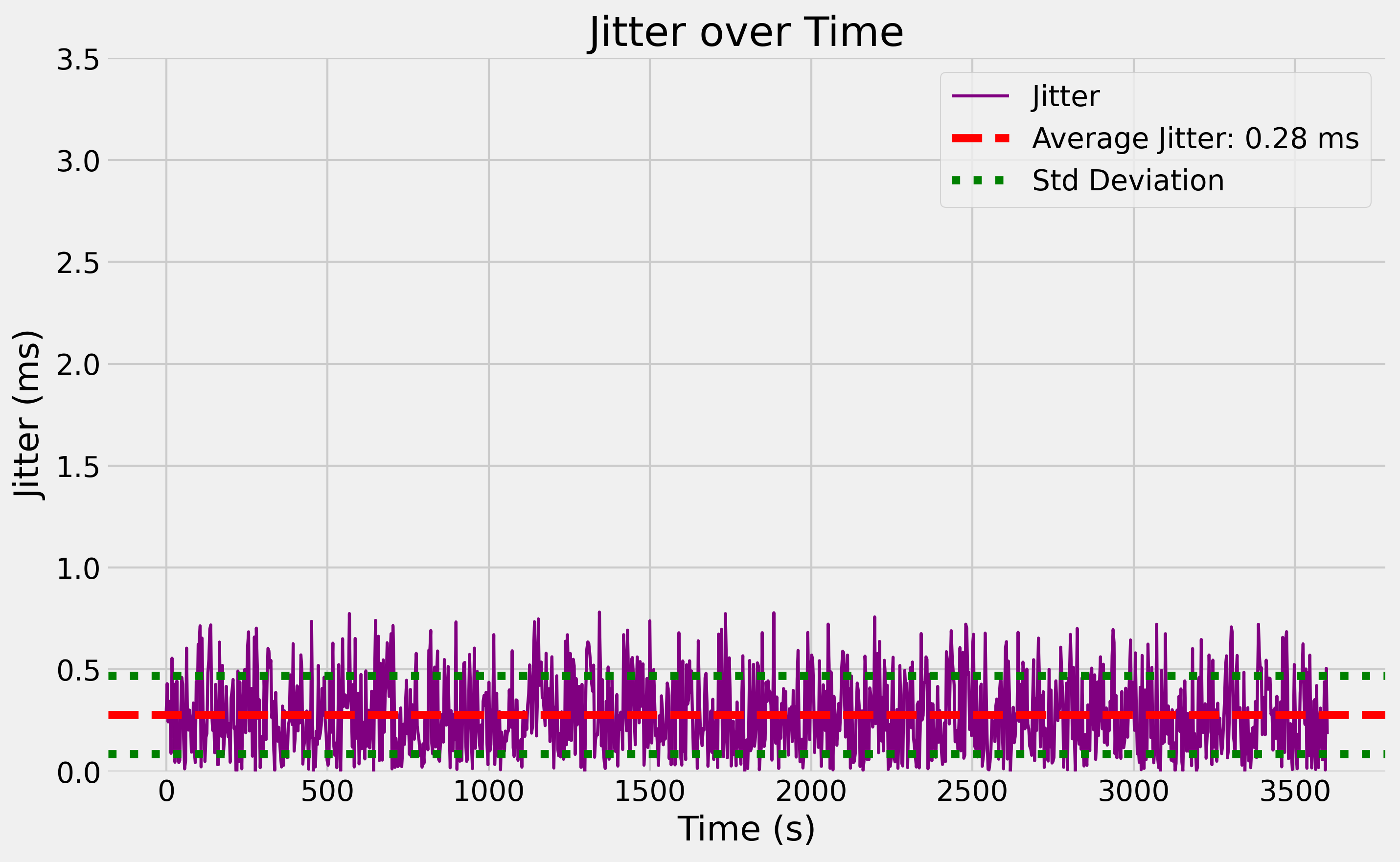}
    }
    \hspace{0.05\textwidth} 
    \subfloat[Decentralized topology (6G)]{
        \includegraphics[width=0.45\textwidth]{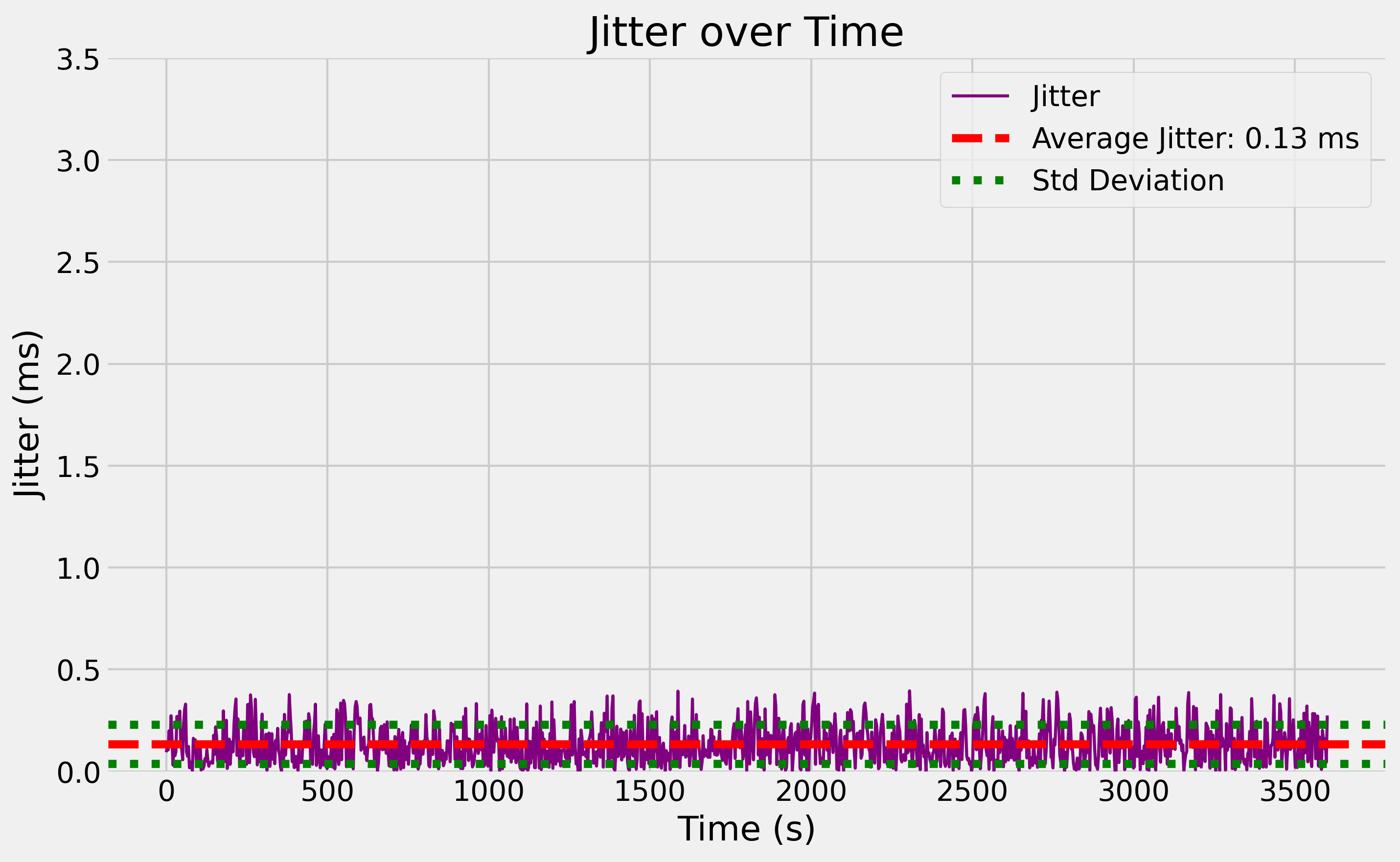}
    }
    \caption{Jitter for all stations.}
    \label{fig:jitter}
\end{figure}

\begin{figure}[h!]
    \centering
    \subfloat[Centralized topology (5G)]{
        \includegraphics[width=0.45\textwidth]{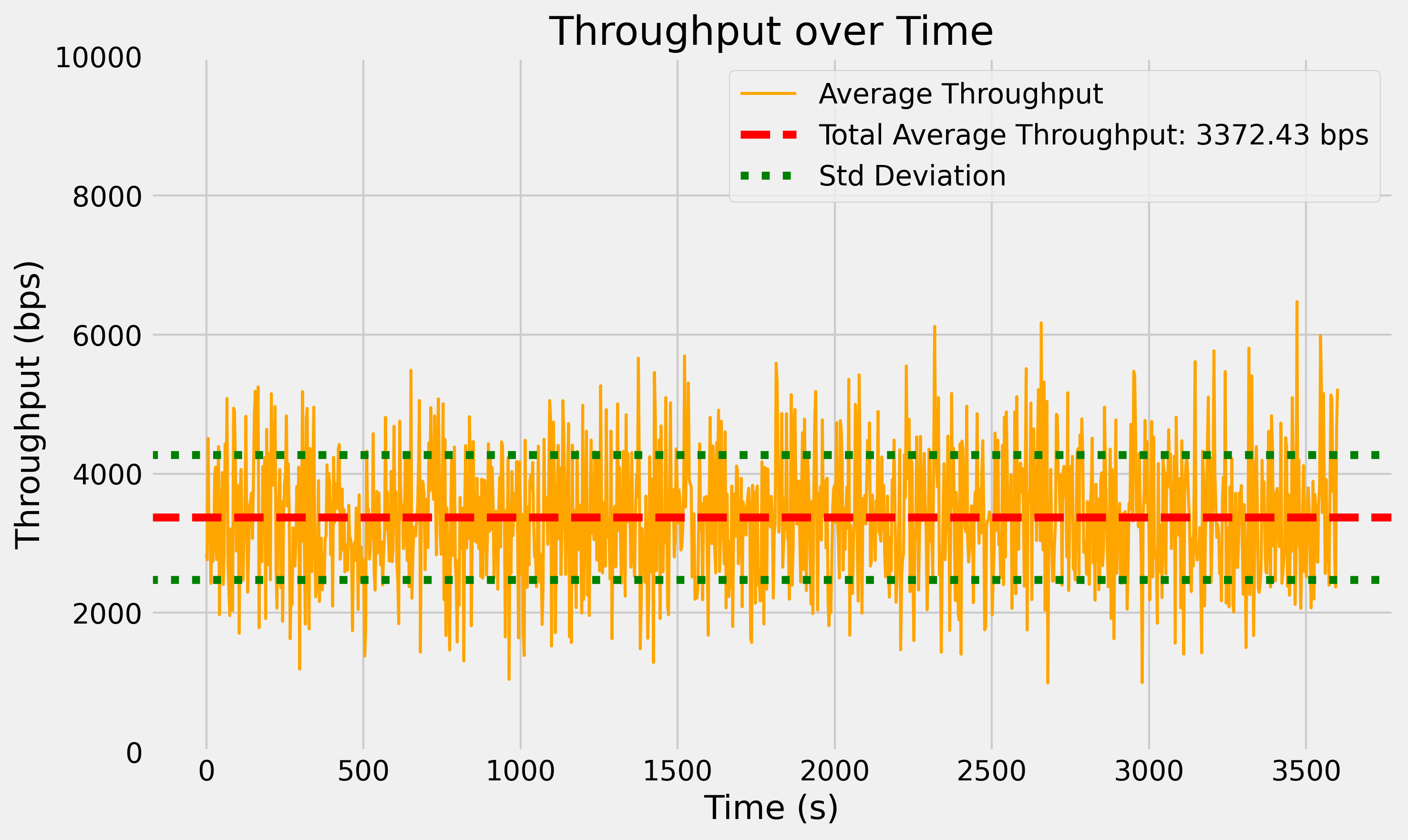}
    }
    \hspace{0.05\textwidth} 
    \subfloat[Decentralized topology (6G)]{
        \includegraphics[width=0.45\textwidth]{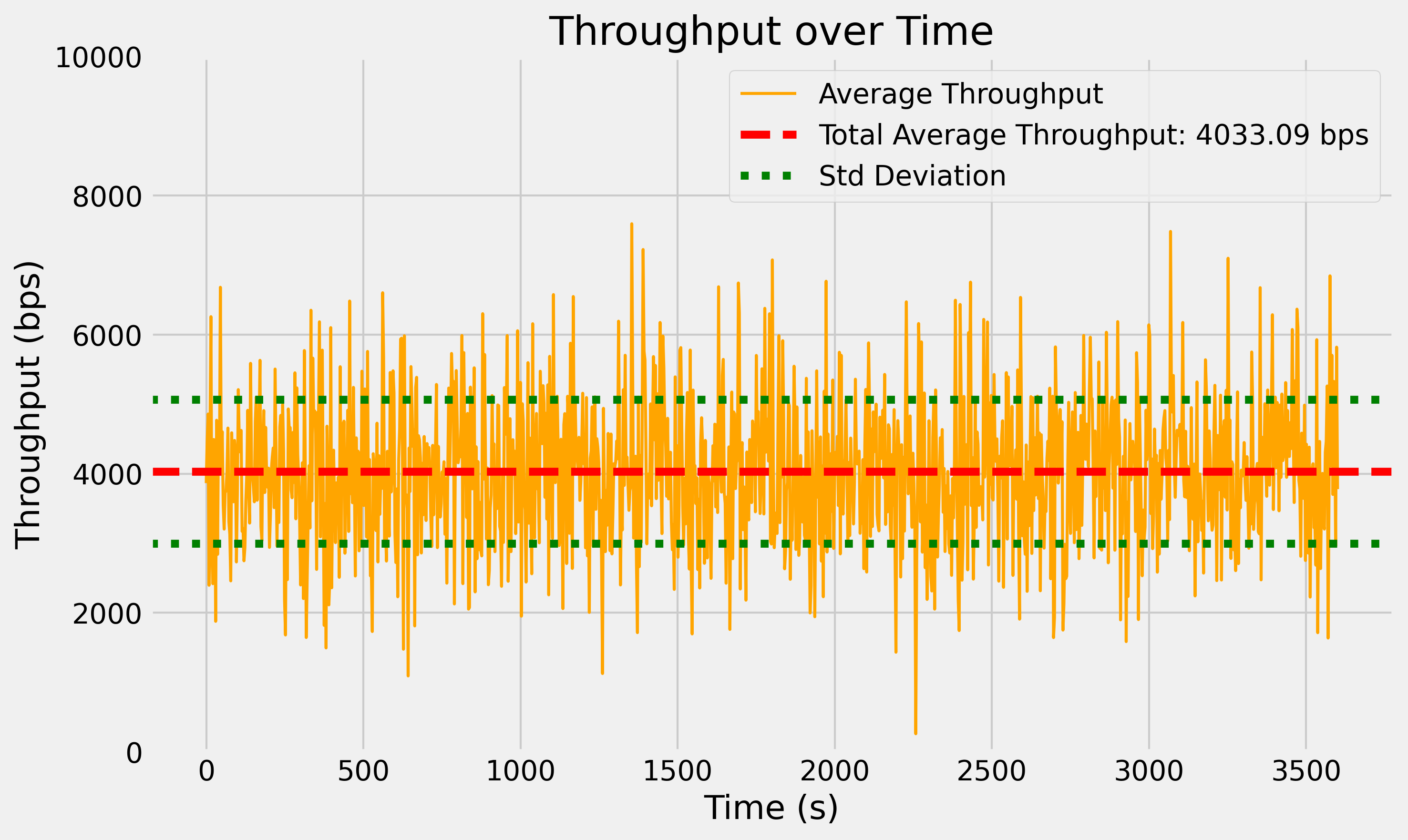}
    }
    \caption{Average throughput for all stations.}
    \label{fig:throughput}
\end{figure}

By comparing the two sets of results obtained in Figures \ref{fig:delay}, \ref{fig:jitter} and \ref{fig:throughput}, it can be observed that the proposed system model approach presents better performance and improvements in terms of low delay and jitter, and high throughput when compared to the centralized topology scenario. Such results can be explained because in the distributed scenario, traffic is spread across multiple clusters, with cluster heads managing the communication between the mobile stations and the host. This decentralized approach can lead to more efficient use of network resources, as each cluster manages its own traffic, reducing congestion on shared paths. The even distribution of traffic load among different cluster heads allows for faster packet processing, resulting in:

\begin{itemize}
    \item \textbf{Average delay was reduced by 16.3\%:} This confirms that distributing routing decisions across multiple servers reduces network congestion and shortens the path length for most packets.
    \item \textbf{Jitter was reduced by 51\%:} A key highlight, indicating that the decentralized system provides more consistent transmission times, which is critical for real-time applications like video streaming, UAV control, or emergency communications.
    \item \textbf{Throughput improved by 15.5\%: } This reflects the system’s enhanced capacity to support high data-rate applications and confirms the effective utilization of network resources enabled by optimal cluster head selection.
\end{itemize}

These improvements can be attributed to multiple factors:

\begin{enumerate}
    \item Local decision-making in clusters, avoiding bottlenecks associated with central servers.
    \item Intelligent cluster head selection, based on proximity and signal strength, which improves intra-cluster communication efficiency.
    \item Machine learning-based prediction of mobility patterns, allowing the network to anticipate and adapt to changes dynamically.
\end{enumerate}

Moreover, in a distributed scenario, each cluster handles an approximately equal amount of traffic, which improves load balancing across the network. Proper load balancing means that no single cluster head is overwhelmed with traffic, preventing bottlenecks. It can also be considered that a distributed clustering approach scales better, as the workload is shared across multiple nodes (cluster heads).

\begin{figure}[h!]
    \centering
    \subfloat[[Total Average Delay]{
        \includegraphics[width=0.45\textwidth]{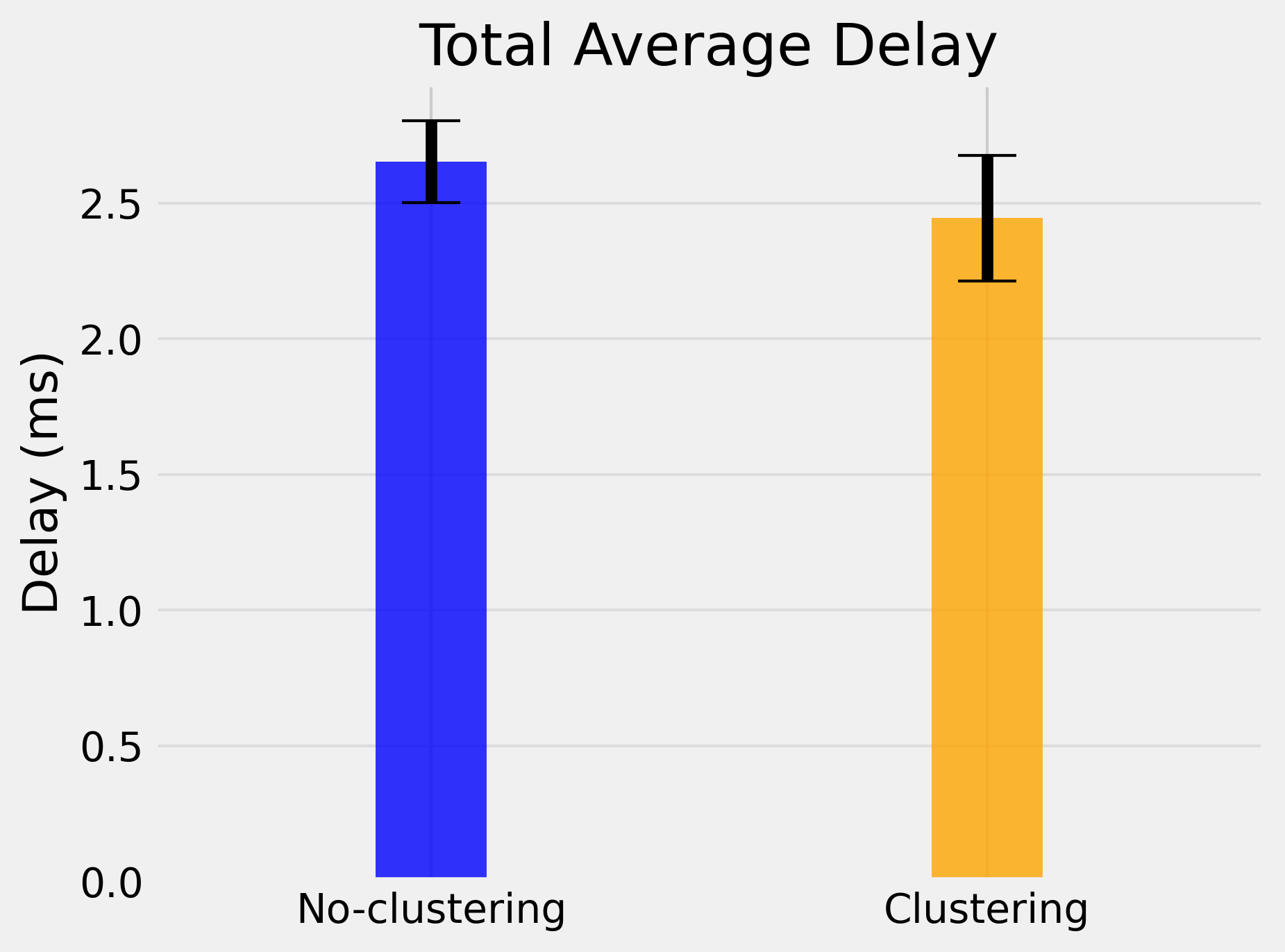}
    }
    \hspace{0.05\textwidth} 
    \subfloat[Total Average Jitter]{
        \includegraphics[width=0.45\textwidth]{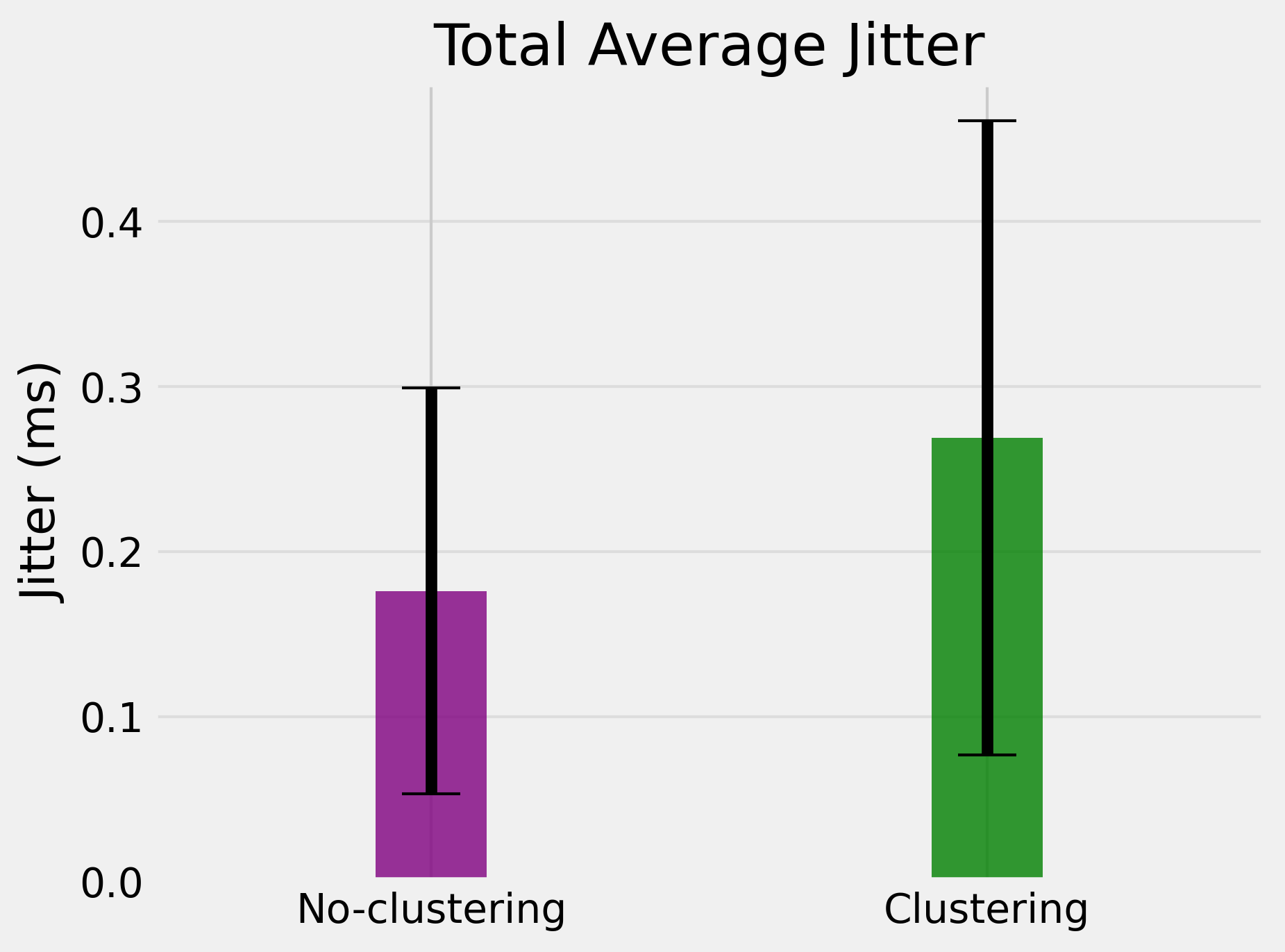}
    }
    \hspace{0.05\textwidth} 
    \subfloat[Average Throughput]{
        \includegraphics[width=0.45\textwidth]{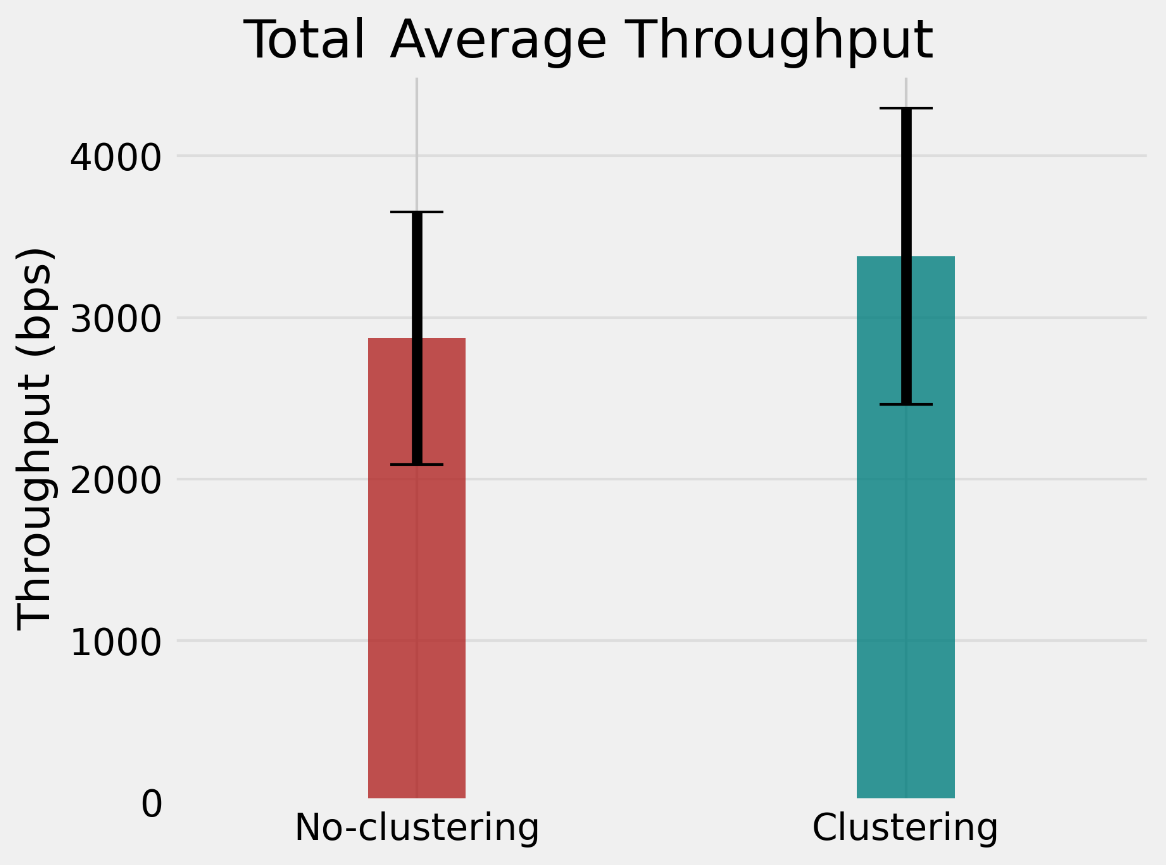}
    }
    \caption{Comparison of metrics results in the centralized Topology (5G) without and with the cluster solution.}
    \label{fig:comp1}
\end{figure}

\begin{figure}[h!]
    \centering
    \subfloat[Total Average Delay]{
        \includegraphics[width=0.45\textwidth]{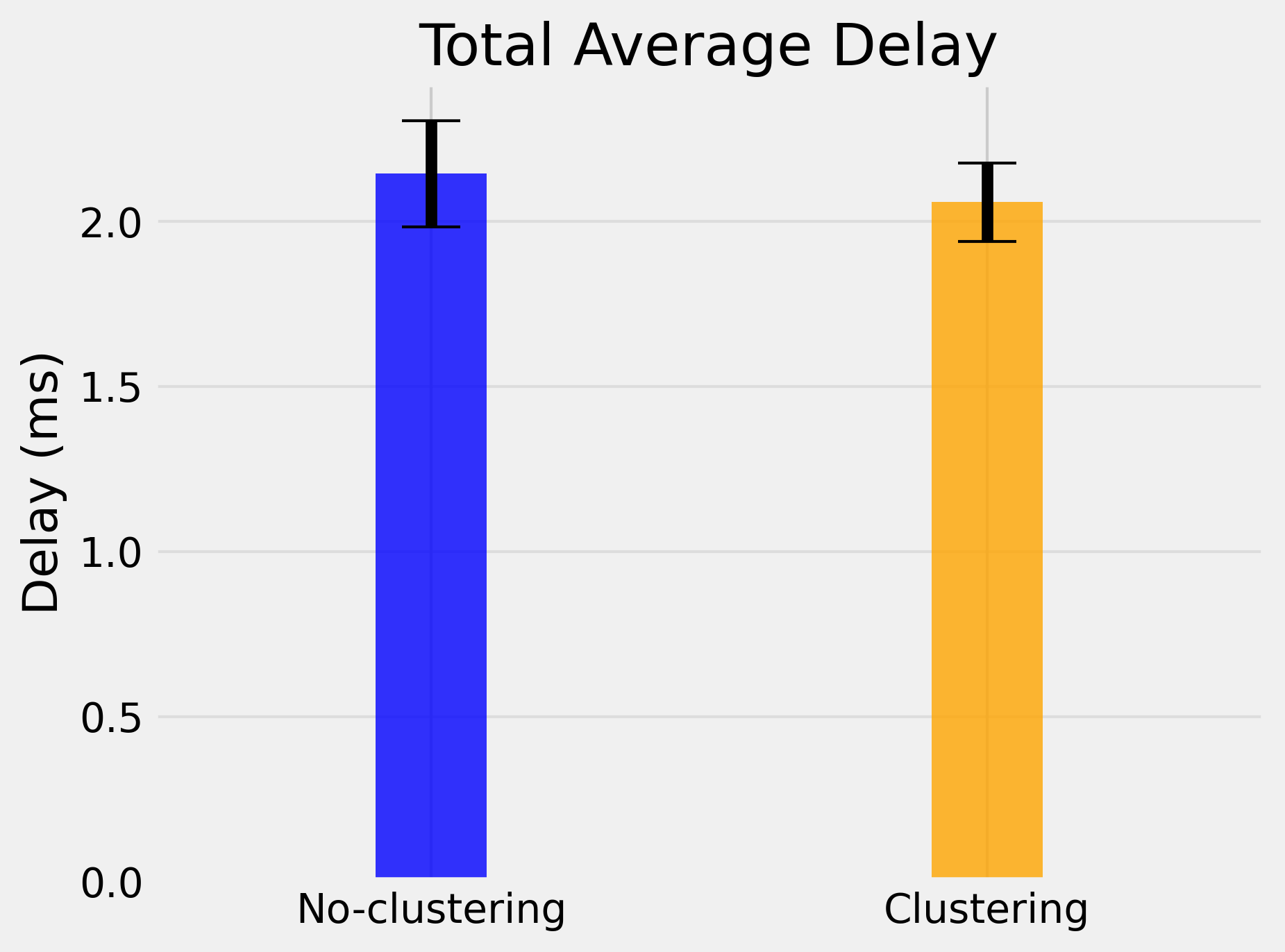}
    }
    \hspace{0.05\textwidth} 
    \subfloat[Total Average Jitter]{
        \includegraphics[width=0.45\textwidth]{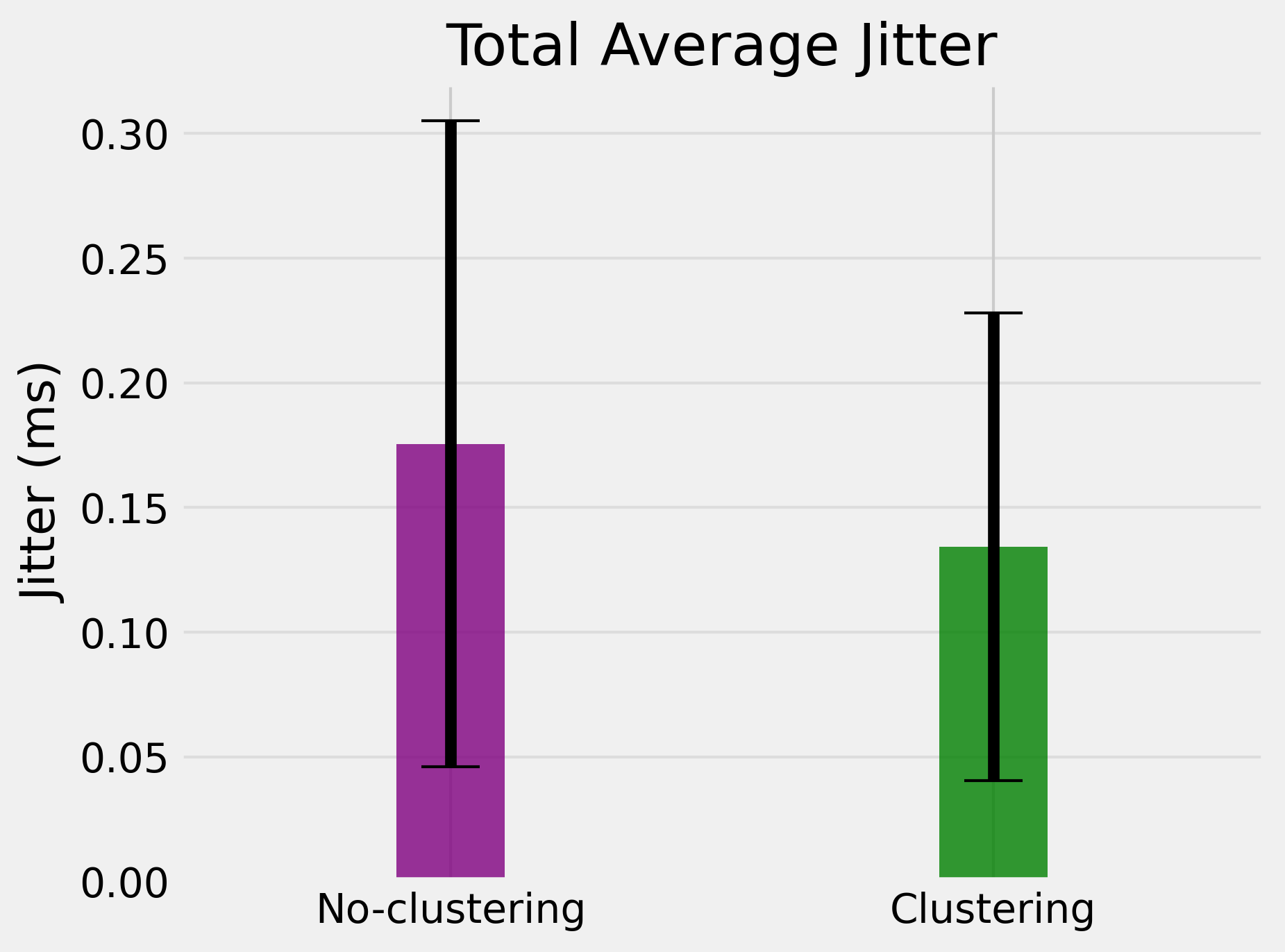}
    }
    \hspace{0.05\textwidth} 
    \subfloat[Total Average Throughput]{
        \includegraphics[width=0.45\textwidth]{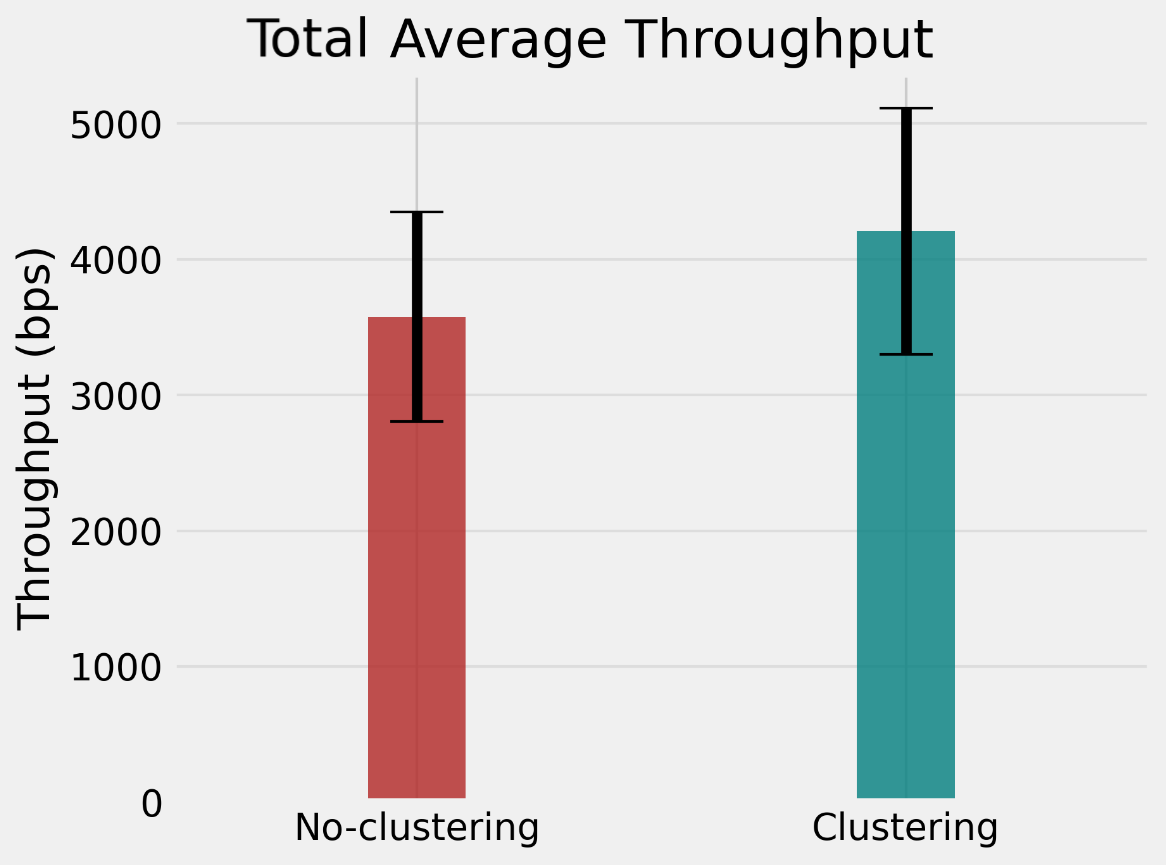}
    }
    \caption{Comparison of metrics results in the decentralized Topology (6G) without and with the cluster solution.}
    \label{fig:comp2}
\end{figure}

In order to evaluate and assess the improvements of the proposed solution in terms of delay, jitter and throughput, the results obtained in the experiments were compared to the results obtained when repeating the experiments, under the same simulations setup, but without using the network clustering. Figures \ref{fig:comp1} and \ref{fig:comp2} present such comparison of results in terms of average and standard deviation of the metrics. In both topologies, clustering significantly outperforms the non-clustering baseline. However, the decentralized approach clearly scales better, particularly in highly dynamic environments like FANETs, validating its relevance for 6G contexts. Whereas all the results demonstrate that the proposed network clustering approach presents benefits in terms of improvement of delay, jitter and throughput for both simulation scenarios, we note that the total average jitter had a small increase when using the network clustering in the centralized topology (5G) scenario.

  {The observed slight increase in jitter under centralized topology can be attributed to the reliance on a single control entity for routing and scheduling, which creates transient queuing delays when multiple UAVs simultaneously forward data. While detailed mitigation is beyond the scope of this study, potential strategies include adaptive scheduling at the central controller, hybrid centralized–distributed clustering to reduce bottlenecks, or predictive load balancing mechanisms to smooth traffic fluctuations. These approaches represent promising directions to reduce jitter in centralized architectures.}

The comparative results summarized in Table~\ref{tab:performance_summary} illustrate the performance impact of the proposed clustering solution in both centralized (5G) and decentralized (6G) network scenarios. Across all evaluated metrics—delay, jitter, and throughput—the clustering approach consistently outperforms the non-clustered baseline. In the 5G topology, clustering reduced average delay by 11.5\% and improved throughput by 9.8\%, although it introduced a slight increase in jitter due to queuing at the cluster heads. In contrast, the 6G decentralized topology exhibited even more substantial gains: delay decreased by 16.3\%, jitter was reduced by 51\%, and throughput increased by 15.5\%. These results confirm that the combination of machine learning-driven mobility prediction and optimized cluster head selection enables more efficient communication, especially in distributed network environments where scalability and low-latency routing are critical.


\begin{table}[h!]
\centering
\caption{Performance Comparison: With vs. Without Clustering}
\begin{tabularx}{\linewidth}{|l|X|X|X|X|}
  ine
\textbf{Metric} & \textbf{5G-No   {clustering}} & \textbf{5G-Clust.} & \textbf{6G-No   {clustering}} & \textbf{6G-Clust.} \\   ine
Delay & High & Mod. ($\downarrow$11.5\%) & Mod. & Low ($\downarrow$18.4\%) \\   ine
Jitter & Mod. & $\uparrow$ Slightly High & Mod. & $\downarrow$ Very Low \\   ine
Throughput & Mod. & $\uparrow$ High (9.8\%) & High & $\uparrow$ Very High (11.7\%) \\   ine
\end{tabularx}
\label{tab:performance_summary}
\end{table}

With these results, it is possible to assume that the proposed system model has benefits in terms of lower delay, lower jitter and higher throughput for network clustering in distributed topologies that are being considered for 6G networks. In summary, the results suggest that ML-driven clustering and optimized cluster head selection are key enablers for achieving the ultra-low latency and high reliability requirements of 6G. Furthermore, the system offers promising levels of scalability and adaptability, setting a foundation for real-world deployments in autonomous UAV networks.

  {It should be emphasized that the objective of this work is not only to propose a new clustering algorithm, but rather to evaluate the impact of integrating machine learning into the clustering and cluster head selection process for UAV networks. For this reason, we use the non-clustered centralized and decentralized topologies as the primary baselines, as these represent the default network configurations without clustering. While direct comparisons with established clustering methods (\textit{e.g.}, LEACH, HEED} \cite{gupta2022proposed},   {or FANET-specific protocols) could provide additional perspective, such evaluations are beyond the scope of this study and are left for future work. This design choice ensures that the observed improvements can be directly attributed to the proposed intelligent clustering framework, rather than generic clustering effects.}

\section{Conclusion}
\label{sec:conclusion}

{This work introduced an intelligent clustering framework for UAV networks tailored to 6G environments, combining XGBoost-based mobility prediction with optimization-driven cluster head selection. The approach effectively reduced delay and jitter while improving throughput, as demonstrated in both centralized (5G) and decentralized (6G) topologies, with the latter offering the highest gains due to distributed routing and load balancing.}

{The proposed design provides a generalizable foundation for future 6G research, with potential extensions toward energy-aware routing, security-enhanced clustering, and edge intelligence for localized data processing. While this study focused on average performance metrics, future work will also examine worst-case and percentile-based delays and jitter to provide stronger reliability guarantees for mission-critical UAV applications. Additional directions include integration with full 5G/6G core simulators, adoption of advanced ML models such as GNNs, Transformers, and federated learning, and multiobjective optimization that jointly considers delay, energy, and link stability. Nevertheless, their deployment in real UAV environments faces challenges including limited onboard resources, communication overhead for distributed training, and robustness to dynamic mobility, which will guide our future extensions. Finally, future work will also complement the theoretical complexity analysis with empirical scalability evaluations, benchmarking execution time across large UAV swarms to ensure computational efficiency at scale.}

\bibliographystyle{unsrt}
\bibliography{references}

\begin{thebibliography}{10}

\bibitem{jpba}
Y.~Jiang, Y.~Zou, H.~Guo, T.~A. Tsiftsis, M.~R. Bhatnagar, R.~C. de~Lamare, and Y.-D. Yao.
\newblock Joint power and bandwidth allocation for energy-efficient heterogeneous cellular networks.
\newblock {\em IEEE Transactions on Communications}, 67(9):6168--6178, 2019.

\bibitem{cappello2023using}
Giorgia~Maria Cappello, Gabriella Colajanni, Patrizia Daniele, Laura Galluccio, Christian Grasso, Giovanni Schembra, and Laura Scrimali.
\newblock Using fanets for 6g cloud-native slice provisioning: A marketplace approach.
\newblock In {\em 2023 IEEE Conference on Network Function Virtualization and Software Defined Networks (NFV-SDN)}, pages 220--226. IEEE, 2023.

\bibitem{pasandideh2023systematic}
Faezeh Pasandideh, Jo{\~a}o Paulo J~da Costa, Rafael Kunst, Wibowo Hardjawana, and Edison~Pignaton de~Freitas.
\newblock A systematic literature review of flying ad hoc networks: State-of-the-art, challenges, and perspectives.
\newblock {\em Journal of Field Robotics}, 40(4):955--979, 2023.

\bibitem{Abdulhae2022ClusterBasedRP}
Omer~T. Abdulhae, Jit~Singh Mandeep, and Mt~Islam.
\newblock Cluster-based routing protocols for flying ad hoc networks (fanets).
\newblock {\em IEEE Access}, 10:32981--33004, 2022.

\bibitem{cbmlrout}
L.~A. L. F.~Da Costa, R.~C. de~Lamare, R.~Kunst, and E.~P. de~Freitas.
\newblock Cluster-based machine learning-driven routing for uav networks in 6g environment.
\newblock {\em IEEE Access}, 13:174957--174969, 2025.

\bibitem{rpreccf}
V.~M.~T. Palhares, A.~R. Flores, and R.~C. de~Lamare.
\newblock Robust mmse precoding and power allocation for cell-free massive mimo systems.
\newblock {\em IEEE Transactions on Vehicular Technology}, 70(5):5115--5120, 2021.

\bibitem{cesg}
S.~Mashdour, R.~C. de~Lamare, and J.~P. S.~H. Lima.
\newblock Enhanced subset greedy multiuser scheduling in clustered cell-free massive mimo systems.
\newblock {\em IEEE Communications Letters}, 27(2):610--614, 2023.

\bibitem{rrs}
A.~R. Flores and R.~C. de~Lamare.
\newblock Robust rate-splitting-based precoding for cell-free mu-mimo systems.
\newblock {\em IEEE Communications Letters}, 29(6):1230--1234, 2025.

\bibitem{rscf}
A.~R. Flores, R.~C. de~Lamare, and K.~V. Mishra.
\newblock Clustered cell-free multi-user multiple-antenna systems with rate-splitting: Precoder design and power allocation.
\newblock {\em IEEE Transactions on Communications}, 71(10):5920--5934, 2023.

\bibitem{rracf}
S.~Mashdour, A.~R. Flores, S.~Salehi, R.~C. de~Lamare, A.~Schmeink, and P.~R.~B. da~Silva.
\newblock Robust resource allocation in cell-free massive mimo systems.
\newblock {\em IEEE Transactions on Communications}, 73(8):5745--5759, 2025.

\bibitem{iddllr}
R.~B.~Di Renna and R.~C. de~Lamare.
\newblock Iterative detection and decoding with log-likelihood ratio based access point selection for cell-free mimo systems.
\newblock {\em IEEE Transactions on Vehicular Technology}, 73(5):7418--7423, 2024.

\bibitem{oclidd}
Tonny Ssettumba, Saeed Mashdour, Lukas T.~N. Landau, Paulo~B. da~Silva, and Rodrigo~C. de~Lamare.
\newblock Iterative interference cancellation for clustered cell-free massive mimo networks.
\newblock {\em IEEE Wireless Communications Letters}, 14(2):509--513, 2025.

\bibitem{Hosseinzadeh2023ANF}
Mehdi Hosseinzadeh, Adil~Hussein Mohammed, Farhan~A. Alenizi, Mazhar~Hussain Malik, Efat Yousefpoor, Mohammad~Sadegh Yousefpoor, Omed~Hassan Ahmed, Amir~Masoud Rahmani, and Lilia Tightiz.
\newblock A novel fuzzy trust-based secure routing scheme in flying ad hoc networks.
\newblock {\em Veh. Commun.}, 44:100665, 2023.

\bibitem{Lnsk2022ReinforcementLR}
Jan L{\'a}nsk{\'y}, Saqib Ali, Amir~Masoud Rahmani, Mohammad~Sadegh Yousefpoor, Efat Yousefpoor, Faheem~Ahmad Khan, and Mehdi Hosseinzadeh.
\newblock Reinforcement learning-based routing protocols in flying ad hoc networks (fanet): A review.
\newblock {\em Mathematics}, 2022.

\bibitem{debasis2023energy}
Kumar Debasis, Lakhan~Dev Sharma, Vijay Bohat, and Robin~Singh Bhadoria.
\newblock An energy-efficient clustering algorithm for maximizing lifetime of wireless sensor networks using machine learning.
\newblock {\em Mobile networks and applications}, 28(2):853--867, 2023.

\bibitem{asaamoning2021dynamic}
Godwin Asaamoning, Paulo Mendes, and Naercio Magaia.
\newblock A dynamic clustering mechanism with load-balancing for flying ad hoc networks.
\newblock {\em IEEE Access}, 9:158574--158586, 2021.

\bibitem{abdulhae2023reinforcement}
Omer~T Abdulhae, Jit~Singh Mandeep, Mohammad~Tariqul Islam, and Md~Shabiul Islam.
\newblock Reinforcement-based clustering in flying ad-hoc networks for serving vertical and horizontal routing.
\newblock {\em IEEE Access}, 11:143881--143895, 2023.

\bibitem{xu2024multi}
Fang Xu, Bin Duo, Yiyuan Xie, Gaofeng Pan, Yandong Yang, Luozhi Zhang, Yichen Ye, Tingnan Bao, Thomas~Aaron Gulliver, and Yuanchen Wang.
\newblock Multi-uav assisted mixed fso/rf communication network for urgent tasks: Fairness oriented design with drl.
\newblock {\em IEEE Transactions on Vehicular Technology}, 2024.

\bibitem{chen2024joint}
Jianrui Chen, Jingjing Wang, Jiaxing Wang, and Lin Bai.
\newblock Joint fairness and efficiency optimization for csma/ca-based multi-user mimo uav ad hoc networks.
\newblock {\em IEEE Journal of Selected Topics in Signal Processing}, 2024.

\bibitem{li2024ground}
Da~Li, Peian Li, Jiabiao Zhao, Jianjian Liang, Jiacheng Liu, Guohao Liu, Yuanshuai Lei, Wenbo Liu, Jianqin Deng, Fuyong Liu, et~al.
\newblock Ground-to-uav sub-terahertz channel measurement and modeling.
\newblock {\em Optics Express}, 32(18):32482--32494, 2024.

\bibitem{xu2025blockchain}
Yueqiang Xu, Haitao Xu, Xuanyu Chen, Heli Zhang, Bin Chen, and Zhu Han.
\newblock Blockchain-based ar offloading in uav-enabled mec networks: A trade-off between energy consumption and rendering latency.
\newblock {\em IEEE Transactions on Vehicular Technology}, 2025.

\bibitem{noman2023machine}
Hafiz Muhammad~Fahad Noman, Effariza Hanafi, Kamarul~Ariffin Noordin, Kaharudin Dimyati, Mhd~Nour Hindia, Atef Abdrabou, and Faizan Qamar.
\newblock Machine learning empowered emerging wireless networks in 6g: Recent advancements, challenges and future trends.
\newblock {\em IEEE Access}, 11:83017--83051, 2023.

\bibitem{sefati2022cluster}
Seyed~Salar Sefati, Simona Halunga, and Roya~Zareh Farkhady.
\newblock Cluster selection for load balancing in flying ad hoc networks using an optimal low-energy adaptive clustering hierarchy based on optimization approach.
\newblock {\em Aircraft Engineering and Aerospace Technology}, 94(8):1344--1356, 2022.

\bibitem{wang2024fast}
Na~Wang, Meng Xiao, Zhongliang Zhao, and Yang Liu.
\newblock A fast weighted clustering algorithm for fanet.
\newblock In {\em 2024 IEEE 99th Vehicular Technology Conference (VTC2024-Spring)}, pages 1--6. IEEE, 2024.

\bibitem{karpagalakshmi2024energy}
RC~Karpagalakshmi, D~Leela Rani, N~Magendiran, and A~Manikandan.
\newblock An energy-efficient bio-inspired mobility-aware cluster p-woa algorithm for intelligent whale optimization and fuzzy-logic-based zonal clustering algorithm in fanet.
\newblock {\em International Journal of Computational Intelligence Systems}, 17(1):258, 2024.

\bibitem{zhang2023capacity}
Haijun Zhang, Miaolin Huang, Huan Zhou, Xianmei Wang, Ning Wang, and Keping Long.
\newblock Capacity maximization in ris-uav networks: A ddqn-based trajectory and phase shift optimization approach.
\newblock {\em IEEE Transactions on Wireless Communications}, 22(4):2583--2597, 2023.

\bibitem{zhou2024fd3pg}
Huan Zhou, Hao Wang, Zhiwen Yu, Guo Bin, Mingjun Xiao, and Jie Wu.
\newblock Federated distributed deep reinforcement learning for recommendation-enabled edge caching.
\newblock {\em IEEE Transactions on Services Computing}, 17(6):3640--3654, 2024.

\bibitem{zhou2025qos}
Huan Zhou, Yadong Lu, Geyong Min, Zhiwen Yu, Liang Wang, Yao Zhang, and Bin Guo.
\newblock Qos-oriented joint resource and trajectory optimization in noma-enhanced uav-mec systems.
\newblock {\em IEEE Transactions on Mobile Computing}, 2025.
\newblock Early Access.

\bibitem{tang2024deep}
Rui Tang, Ruizhi Zhang, Yongjun Xu, and Chau Yuen.
\newblock Deep reinforcement learning-based resource allocation for multi-uav-assisted full-duplex wireless-powered iot networks.
\newblock {\em IEEE Transactions on Cognitive Communications and Networking}, 2024.

\bibitem{prakash2024reinforcement}
M~Prakash, S~Neelakandan, and Bong-Hyun Kim.
\newblock Reinforcement learning-based multidimensional perception and energy awareness optimized link state routing for flying ad-hoc networks.
\newblock {\em Mobile Networks and Applications}, 29(2):315--333, 2024.

\bibitem{jaiswal2024comparative}
Khushbu Jaiswal, Sudesh Kumar, and Neeraj~Kumar Rathore.
\newblock Comparative analysis of machine learning application on routing protocols in fanet.
\newblock In {\em 2024 13th International Conference on System Modeling \& Advancement in Research Trends (SMART)}, pages 706--711. IEEE, 2024.

\bibitem{li2024reinforcement}
Jieling Li, Liang Xiao, Xuchen Qi, Zefang Lv, Qiaoxin Chen, and Yong-Jin Liu.
\newblock Reinforcement learning based energy-efficient fast routing for fanets.
\newblock {\em IEEE Transactions on Communications}, 2024.

\bibitem{song2025gnnppor}
Jian Song, Jing Li, Qingwang Wang, Yebo Gu, and Tao Shen.
\newblock Gnnppor: A proximal policy optimization multi-factor joint routing approach based on graph neural networks in fanets.
\newblock {\em IEEE Networking Letters}, 2025.

\bibitem{priyadharshini2024efficient}
SP~Priyadharshini and P~Balamurugan.
\newblock An efficient ddos attack detection and prevention model using fusion heuristic enhancement of deep learning approach in fanet sector.
\newblock {\em Applied Soft Computing}, 167:112438, 2024.

\bibitem{luo2025convergence}
Haoxiang Luo, Gang Sun, Cheng Chi, Hongfang Yu, and Mohsen Guizani.
\newblock Convergence of symbiotic communications and blockchain for sustainable and trustworthy 6g wireless networks.
\newblock {\em IEEE Wireless Communications}, 32(2):18--25, 2025.

\bibitem{9286738}
Gang Sun, Zhu Xu, Hongfang Yu, and Victor Chang.
\newblock Dynamic network function provisioning to enable network in box for industrial applications.
\newblock {\em IEEE Transactions on Industrial Informatics}, 17(10):7155--7164, 2021.

\bibitem{10714036}
Gang Sun, Yuhui Wang, Hongfang Yu, and Mohsen Guizani.
\newblock Proportional fairness-aware task scheduling in space-air-ground integrated networks.
\newblock {\em IEEE Transactions on Services Computing}, 17(6):4125--4137, 2024.

\bibitem{chen2024and}
Peipei Chen, Lailong Luo, Deke Guo, Guoming Tang, Baokang Zhao, Yan Li, and Xueshan Luo.
\newblock Why and how lasagna works: a new design of air-ground integrated infrastructure.
\newblock {\em IEEE Network}, 38(2):132--140, 2024.

\bibitem{Chen2016XGBoostAS}
Tianqi Chen and Carlos Guestrin.
\newblock Xgboost: A scalable tree boosting system.
\newblock {\em Proceedings of the 22nd ACM SIGKDD International Conference on Knowledge Discovery and Data Mining}, 2016.

\bibitem{syakur2018integration}
Muhammad~Ali Syakur, B~Khusnul Khotimah, EMS Rochman, and Budi~Dwi Satoto.
\newblock Integration k-means clustering method and elbow method for identification of the best customer profile cluster.
\newblock In {\em IOP conference series: materials science and engineering}, volume 336, page 012017. IOP Publishing, 2018.

\bibitem{zhao2008knee}
Qinpei Zhao, Ville Hautamaki, and Pasi Fr{\"a}nti.
\newblock Knee point detection in bic for detecting the number of clusters.
\newblock In {\em International conference on advanced concepts for intelligent vision systems}, pages 664--673. Springer, 2008.

\bibitem{Costa2021QFANETIQ}
Luis Antonio Leite~Francisco da~Costa, Rafael Kunst, and Edison~Pignaton de~Freitas.
\newblock Q-fanet: Improved q-learning based routing protocol for fanets.
\newblock {\em Comput. Networks}, 198:108379, 2021.

\bibitem{daCosta2024RoutingPE}
Luis Antonio L.~F. da~Costa, Rafael Kunst, Rodrigo~C. de~Lamare, and Edison~Pignaton de~Freitas.
\newblock Routing performance evaluation in centralized and distributed fanets for 6g networks.
\newblock {\em 2024 19th International Symposium on Wireless Communication Systems (ISWCS)}, pages 1--6, 2024.

\bibitem{Chai2014RootMS}
Tianfeng Chai and Roland~R. Draxler.
\newblock Root mean square error (rmse) or mean absolute error (mae)? – arguments against avoiding rmse in the literature.
\newblock {\em Geoscientific Model Development}, 7:1247--1250, 2014.

\bibitem{alba2021enabling}
Alberto~Mart{\'\i}nez Alba, Shakthivelu Janardhanan, and Wolfgang Kellerer.
\newblock Enabling dynamically centralized ran architectures in 5g and beyond.
\newblock {\em IEEE Transactions on Network and Service Management}, 18(3):3509--3526, 2021.

\bibitem{spa}
R.~C. De~Lamare and R.~Sampaio-Neto.
\newblock Minimum mean-squared error iterative successive parallel arbitrated decision feedback detectors for ds-cdma systems.
\newblock {\em IEEE Transactions on Communications}, 56(5):778--789, 2008.

\bibitem{jidf}
R.~C. de~Lamare and R.~Sampaio-Neto.
\newblock Adaptive reduced-rank processing based on joint and iterative interpolation, decimation, and filtering.
\newblock {\em IEEE Transactions on Signal Processing}, 57(7):2503--2514, 2009.

\bibitem{rsprec}
A.~R. Flores, R.~C. de~Lamare, and B.~Clerckx.
\newblock Linear precoding and stream combining for rate splitting in multiuser mimo systems.
\newblock {\em IEEE Communications Letters}, 24(4):890--894, 2020.

\bibitem{lrcc}
H.~Ruan and R.~C. de~Lamare.
\newblock Distributed robust beamforming based on low-rank and cross-correlation techniques: Design and analysis.
\newblock {\em IEEE Transactions on Signal Processing}, 67(24):6411--6423, 2019.

\bibitem{zeebaree2017combination}
Diyar~Qader Zeebaree, Habibollah Haron, Adnan~Mohsin Abdulazeez, and SR~Zeebaree.
\newblock Combination of k-means clustering with genetic algorithm: A review.
\newblock {\em International Journal of Applied Engineering Research}, 12(24):14238--14245, 2017.

\bibitem{rana2011review}
Sandeep Rana, Sanjay Jasola, and Rajesh Kumar.
\newblock A review on particle swarm optimization algorithms and their applications to data clustering.
\newblock {\em Artificial Intelligence Review}, 35:211--222, 2011.

\bibitem{Cui2022}
Yanpeng Cui, Qixun Zhang, Zhiyong Feng, Zhiqing Wei, Ce~Shi, and Heng Yang.
\newblock Topology-aware resilient routing protocol for fanets: An adaptive q-learning approach.
\newblock {\em IEEE Internet of Things Journal}, 9(19):18632--18649, 2022.

\bibitem{Fontes2015MininetWiFiES}
Ramon dos Reis~Fontes, Samira Afzal, Samuel Henrique~Bucke Brito, Mateus A.~S. Santos, and Christian~Esteve Rothenberg.
\newblock Mininet-wifi: Emulating software-defined wireless networks.
\newblock {\em 2015 11th International Conference on Network and Service Management (CNSM)}, pages 384--389, 2015.

\bibitem{henderson2008network}
Thomas~R Henderson, Mathieu Lacage, George~F Riley, Craig Dowell, and Joseph Kopena.
\newblock Network simulations with the ns-3 simulator.
\newblock {\em SIGCOMM demonstration}, 14(14):527, 2008.

\bibitem{3GPPTraf68:online}
3gpp xr-01-xr traffic model and kpi.
\newblock \url{https://shorturl.at/vIzZm}.
\newblock (Accessed on 09/02/2024).

\bibitem{926982}
W.R. Heinzelman, A.~Chandrakasan, and H.~Balakrishnan.
\newblock Energy-efficient communication protocol for wireless microsensor networks.
\newblock In {\em Proceedings of the 33rd Annual Hawaii International Conference on System Sciences}, pages 10 pp. vol.2--, 2000.

\bibitem{gupta2022proposed}
Ayan~Das Gupta, K~Sathiyasekar, R~Krishnamoorthy, S~Arun, R~Thiyagarajan, and S~Padmapriya.
\newblock Proposed ga algorithm with h-heed protocol for network optimization using machine learning in wireless sensor networks.
\newblock In {\em 2022 Second International Conference on Artificial Intelligence and Smart Energy (ICAIS)}, pages 1402--1408. IEEE, 2022.

\end{thebibliography}

\end{document}